# Internal and External Field Effects upon Crystal Field Excitations in $REFeO_3$ (RE = $Nd^{3+}$, $Er^{3+}$, $Yb^{3+}$, $Pr^{3+}$, and $Ho^{3+}$)

Guochu Deng*

*Australian Centre for Neutron Scattering, Australian Nuclear Science and Technology Organisation, New Illawarra Road, Lucas Heights, NSW 2234, Australia*

## Abstract

We systematically investigated crystal-field (CF) excitations of Kramers ($Nd^{3+}$, $Er^{3+}$, $Yb^{3+}$) and non-Kramers ($Pr^{3+}$, $Ho^{3+}$) rare-earth ions in $REFeO_3$ using optimized CF simulations. Internal magnetic fields from the $Fe^{3+}$ and $RE^{3+}$ sublattices split the ground-state doublets of all Kramers ions, generating low-energy excitations around 1 meV. In non-Kramers systems, low-energy excitations arise only when the ground state forms an accidental pseudo-doublet, as observed for $Ho^{3+}$ in $HoFeO_3$; such pseudo-doublets exhibit field-induced splitting analogous to Kramers ions. In contrast, true singlet ground states, exemplified by $Pr^{3+}$, show no zero-field splitting. Strong anisotropies are found in both internal- and external-field responses of the CF excitations in these $REFeO_3$. These results provide a unified explanation for the anomalous Zeeman splitting of CF ground states in $REFeO_3$.

## 1. Introduction

Rare earth transition metal oxides (RE-TMOs) are fascinating due to their combination of strong electron correlations, complex magnetism, exotic ground states, and intricate coupling phenomena.[1-4] This complexity gives rise to a broad range of fundamental scientific questions and potential applications, making these materials a vibrant area of research in condensed matter physics and materials science. Exotic phenomena such as high-temperature superconductivity,[3] frustrated quantum magnetism,[5,6] multiferroicity,[7] and topological properties (e.g., insulators[4] and semimetals) have been discovered in RE-TMOs. These unique properties make RE-TMOs highly interesting both for fundamental scientific research and a wide range of technological applications, from the discovery of quantum spin liquids (QSLs) to the development of spintronic and multiferroic devices. Numerous studies have shown that the rich physics of RE-TMOs are strongly linked to the magnetism from both the rare earth and transition metal sites,[8,9] strong spin-orbit coupling,[4,10] crystal field (CF) effects,[8,9,11,12] significant magnetic anisotropy,[13-15] electron-phonon coupling,[16] and complex magnetic competition.[4,10,17,18]

CF excitations are widely observed and studied phenomena in RE-TMOs. For rare earth ions incorporated into a crystal lattice, the surrounding electric fields from neighbouring ligands split the energy levels of their 4*f* orbitals. Transitions between these split levels, namely, CF excitations, can be measured using inelastic neutron scattering and Raman spectroscopy. Such excitations can be modelled using Stevens operator-equivalent formalism, combined with symmetric analysis under the single-ion approximation. Many studies have demonstrated that CF effects play critical roles in suppressing superconductivity,[19] enhancing magnetic anisotropy,[20] and inducing multiferroicity,[21] and many other effects. Thus, studying CF excitations in RE-TMOs is essential for gaining a deeper understanding of their magnetism.

* Corresponding Author, Email: guochu.deng@ansto.gov.au

Rare earth orthoferrites ($REFeO_3$) are a series of RE-TMOs with the space group *Pbmn*, which have long been of interest due to their complex magnetic properties, such as spin reorientation transition. Among these, CF excitations of rare earth ions have been a focal point of study. Recently, an intriguing low-energy CF excitation (< 1 meV) was observed in $ErFeO_3$, showing a strong temperature dependence—an unusual phenomenon that has rarely been reported. B. Z. Malkin et al.[22] reported that the Yb, Er, Sm-doped $Cs_2NaYF_6$ demonstrate a common splitting feature in the ultrafine CF excitation spectra, which was attributed to the random strain and lattice deformation. However, lattice deformations could not be used to simply explain the observed energy shift of the low-energy CF excitations in $REFeO_3$ because inelastic neutron scattering experiments clearly show the correlation between the energy shift of CF excitation and the formation of long-range magnetic order of rare earth ions. Therefore, the origin and mechanism behind this temperature-dependent behaviour remain open questions, meriting further investigation.

Several other $REFeO_3$ compounds also exhibit low-energy CF excitations although temperature dependencies have not been extensively documented. For instance, $NdFeO_3$ displays a CF excitation at 0.8 meV at 1.5 K,[23] while $YbFeO_3$ shows a CF excitation below 1 meV at 1 K,[8] which splits into two peaks below and above 1 meV at 1.5 K. Low-energy CF excitation in $ErFeO_3$ show strong temperature dependency.[11] Similarly, recent studies have reported a low-energy excitation in $HoFeO_3$.[24,25] Whether these occurrences are coincidental or share a universal underlying mechanism remains unclear. Importantly, these low-energy excitations are observed in both Kramers ($Er^{3+}$, $Nd^{3+}$, and $Yb^{3+}$) and non-Kramers ($Ho^{3+}$) ions. This raises questions about whether these excitations are inherently linked to both types of ions and what role the $C_s$ point group symmetry at the rare earth 4*c* site might play. Such low-energy CF excitations are rare in other rare earth compounds, highlighting the need to unravel these mysteries. Answering these questions could significantly enhance our understanding of CF effects in the entire $REFeO_3$ series.

In this study, we systematically investigate the CF excitations of both Kramers and non-Kramers rare earth ions ($Er^{3+}$, $Yb^{3+}$, $Pr^{3+}$, and $Ho^{3+}$) in $REFeO_3$ using the Stevens operator-equivalent formalism to model reported CF excitation data. We explore the effects of magnetic fields by considering internal magnetic contributions and applying external fields based on the established CF model. Our results reveal that internal magnetic fields are prevalent in $REFeO_3$ and significantly influence CF excitation energies. Interestingly, Kramers and non-Kramers ions exhibit distinct behaviours under these magnetic fields. In Kramers ion orthoferrites, low-energy excitations stem from ground-state doublet splitting induced by internal magnetic fields and the low local symmetry $C_s$, suggesting that such excitations are intrinsic to all Kramers ions. Conversely, non-Kramers ions typically do not show low-energy CF excitations because their CF levels are non-degenerate singlets, which remain unaffected by external fields. However, low-energy excitations can still occur in non-Kramers $REFeO_3$ when magnetic fields (internal or external) break symmetry, reviving otherwise suppressed singlet transitions, like the case of $HoFeO_3$. These insights highlight the intricate interplay between CF effects and magnetic interactions, deepening our understanding of these complex materials.

## 2. Experiment and Computation

The 4*f* electrons of rare-earth elements exhibit behaviours that are markedly different from those of the 3d electrons in transition metals. In rare-earth ions, the 4*f* electrons are highly localized because they are strongly shielded by the outer 5*s* and 5*p* electron shells. Consequently, spin–orbit coupling (SOC) in 4*f* systems is very strong, and the total angular momentum *J* remains a good quantum number. In contrast, SOC in *3d* transition metals is relatively weak and is often treated as a perturbation, with the orbital angular momentum largely quenched by the CF effect.

For a free rare-earth ion, the strong SOC splits the 4*f* electronic configuration into a series of well-defined multiplets characterized by different *J* values. These multiplets are separated by energies on the order of hundreds of meV. Owing to the strong shielding of the 4*f* electrons, the electrostatic CF generated by the surrounding ligands is comparatively weak and does not significantly mix different J multiplets. Consequently, the CF interaction is usually treated as a perturbation acting within a given SOC-split multiplet. The CF lifts the degeneracy of the (2*J*+1)-fold ground-state multiplet, producing a set of CF-split sublevels. In most practical cases, only the CF splitting of the ground-state multiplet is considered. The resulting CF excitation energies typically lie in the range from a few meV to several tens of meV, and only rarely approach the hundred-meV scale.

Due to the localized properties of 4*f* electrons in rare earth elements, the exchange interactions between 4*f* electrons (or even with other magnetic ions) are very weak. The low ordering temperature or lack of long-range magnetic order in many rare earth compounds provide a solid support to this conclusion. Even though the 4*f* electron could couple to the lattice vibration, such as phonon, the impacts are usually also small. Generally, thus, it is safe to introduce the single-ion approximation when modelling the CF splitting theoretically. The CF splitting in rare earth elements can be described by the Stevens operator-equivalent method. Considering the Stevens operator equivalents and the Zeeman effect from magnetic fields, the Hamiltonian can be written in the following way.

$$\boldsymbol{H_{CF}} = \sum_{\boldsymbol{k,q}} \boldsymbol{B_k^q O_k^q} + \boldsymbol{g_J \mu_B B J}$$

where $B_k^q$ are the CF parameters, $O_k^q$ are the Stevens operator-equivalents which is in the form of a polynomial of $J_z$, $J_+$, $J_-$, $g_J$ is the Landé *g*-factor for the specific ion's ground state, $\mu_B$ is the Bohr magneton, $\boldsymbol{B}$ is the magnetic field. The CF parameters are dependent on the multiplet itself. Most studies focus on the splitting of the ground state multiplet since most of rare earth elements stays at the ground state at room temperature and below. It is worthwhile to point out that the $B_k^q$ parameters are highly dependent on the local symmetry of the rare earth ions in the lattice. For example, for the high symmetry group $O_h$, there are only $B_4^0$, $B_4^4$, $B_6^0$, and $B_6^4$. In the case of $REFeO_3$, the RE ions occupy the 4*c* sites which has $C_s$ symmetry group, which results in 15 $B_k^q$ parameters.

The CF models for Kramers ($NdFeO_3$, $ErFeO_3$, $YbFeO_3$) and non-Kramers ($PrFeO_3$, $HoFeO_3$) rare earth orthoferrites were constructed using the Stevens operator-equivalent method, fitted to previously reported CF peaks for each compound. These models were employed to compute eigenstates, excited state peak positions, and intensities at base temperature for each compound. To investigate magnetic field effects, internal and external magnetic fields along the *x*-, *y*-, and *z*-directions were incorporated into the model, with the simulated results being analysed and presented. Additionally, the impact of supplementary external magnetic fields, being applied along the same directions as an assumed internal field of 10 T along the three main axes, was examined to assess their influence on existing CF excitation peaks. The dependence of peak positions and intensities on external magnetic fields was also evaluated. Model fitting to experimental data was performed using a custom Python package, developed by the author, which integrates the Mantid CF Application Programming Interface (API)[26] as the core framework for modelling and calculations. Some results were double checked by using the other Python package PyCrystalField.

### 3. Results and discussions

In $REFeO_3$, the rare earth ions RE occupied the 4*c* site of the space group *Pbmn*. This site has a Cs local double group. According to the group theory analysis, the double group $C_s$ has four irreducible representations: two non-Kramers ($\Gamma_1$, $\Gamma_2$) and two Kramers ($\Gamma_3$, $\Gamma_4$), see the character table in Table S1 in the supplementary materials. In groups with very low symmetry, like $C_s$, all Kramers doublets belong to a single, two-dimensional irreducible representation. Thus, the Kramers rare earth ions such as $Nb^{3+}$, $Er^{3+}$, $Yb^{3+}$ have the $\Gamma_3$, $\Gamma_4$ irreducible representations while the non-Kramers rare earth ions such as $Pr^{3+}$ and $Ho^{3+}$ have the $\Gamma_1$ or $\Gamma_2$ irreducible representations. It is worthwhile to notice that the two 2D irreducible representation $\Gamma_3$ and $\Gamma_4$ are identical. The identical $\Gamma_3$ and $\Gamma_4$ means that all Kramers doublets must belong to the same single, two-dimensional irreducible representation. The main reason behind this is that the $C_s$ symmetry is very low only having the identity E and one reflection $\sigma_h$ symmetry operation. The energy levels of any Kramers ion must come in degenerate pairs, which are represented by a two-dimensional irreducible representation. This is the secondary reason for the identical $\Gamma_3$ and $\Gamma_4$. In the following sections, we will discuss the internal and external magnetic field impacts to the CF excitations of both Kramers and non-Kramers rare earth ions in $REFeO_3$.

### 3.1 *Kramers Ions $Nd^{3+}$, $Er^{3+}$ and $Yd^{3+}$ in $REFeO_3$*

Kramers' degeneracy theorem states that rare earth ions with half-integer spin exhibit double degeneracies at all CF levels due to time-reversal symmetry. Given the total spin in the 4*f* orbitals of $Nd^{3+}$, $Er^{3+}$, and $Yb^{3+}$ in $REFeO_3$, the CF energy levels of these Kramers ions should consist of degenerate doublets or, depending on local symmetry, quartets, with no singlets expected. Consequently, we anticipate shared characteristics in the CF excitation spectra of these ions in $REFeO_3$. In the following sections, we will examine the CF excitation spectra of $Nd^{3+}$, $Er^{3+}$, and $Yb^{3+}$ individually, analysing the effects of internal and external magnetic fields on these spectra.

#### 3.1.1 $NdFeO_3$

The CF excitations of $NdFeO_3$ were carefully studied at different temperatures within different energy scales using inelastic neutron scattering technique by M. Loewenhaupt et al.[27] and R. Przenioslo et al.[28] Five CF excitation peaks were observed and reported in Ref [28]. Their peak positions and relative intensities are listed in Table 1. Additionally, one peak in the low-energy range (< 1 meV) was also observed and demonstrated certain temperature dependency. This peak is very similar to the low-energy excitation observed in $ErFeO_3$ at low temperature.[11] According to the analysis in our previous $ErFeO_3$ work, the low energy excitation in $ErFeO_3$ should be attributed to the ground state CF excitation, which was shifted by the internal magnetic fields from the $Fe^{3+}$ and $Er^{3+}$ sublattices. Similarly, we speculate that the low-energy excitation observed in $NdFeO_3$ comes from the ground state CF excitation of $Nd^{3+}$ and is shifted by the internal magnetic fields. However, no further analysis was conducted by the authors to explain how this happened in $NdFeO_3$ in detail. Our recent work has carefully explained and simulated the CF excitation energy levels evolving with the internal and external magnetic fields in $ErFeO_3$.[11] It is highly interesting to carry out similar analyses to the CF excitations in $NdFeO_3$.

**Table 1.** CF excitation peaks and intensities reported for $NdFeO_3$ [2]

| | | | | |
|---|---|---|---|---|
| Energy [meV] | 10.4 ± 0.1 | 22.7 ± 0.6 | 45.4 ± 0.8 | 60.8 ± 1.4 |
| Intensity [a.u.] | 100.0 ± 4.0 | 23.0 ± 4.2 | 41.1 ± 5.5 | 11.6 ± 3.4 |

With the reported CF excitation energies and intensities in Table 1, we build a CF model based on the local point group $C_s$ with the initial parameters calculated from the point-charge model. Further fitting the CF model of $Nd^{3+}$ to the experimental peaks in Table 1 generates the fitted CF parameters shown in Table 2. The internal and external magnetic fields have impacts on the CF excitation energies. With the parameters in Table 2, the internal and external magnetic field effects have been studied through simulations.

The internal fields along the *x*, *y*, and *z* directions were simulated and plotted in Figs. 1, 2, and 3, respectively. As shown in these figures, both the internal and external magnetic fields have strong impacts to the $Nd^{3+}$ CF excitation energies. Almost all the excitation peaks shifted and split under the internal fields. Figs. 1 and 2 clearly show that the internal fields along the *x* and *y* directions split the ground state and drive the ground state excitation to higher energies in the range from 0 to 2 meV. Such energy shifts agree well with the observed low-energy excitation below 1 meV in $NdFeO_3$ by R. Przenioslo et al.[2] The other peak at ~ 22 meV undergoes a similar splitting and shifting effect as well, while the peak at ~ 10 meV seems no significant change. When the internal fields are along the *z* direction, as shown in Fig. 3, the ground state CF peak shows an energy shift to higher energy with a larger weight distribution than the shifts in the *x* and *y* fields. However, both the second peak at ~ 10 meV and the third peak at ~ 22 meV also show splitting and shifts in the *z* fields while the peak at ~ 45.5 meV almost shows no change.

The effects of external magnetic fields along the *x*, *y*, and *z* directions were simulated and presented in Figs. S1, S2, and S3 of the Supplemental Material. The external fields influence individual excitation peaks in a manner similar to internal fields but induce more significant energy shifts.

The effects of external magnetic fields on CF excitation spectra were simulated with an internal field of 10 T along all directions. Figs. 4 and 5 show the results for positive and negative *y*-direction fields, respectively. Negative fields monotonically increase the excitation energy of the split ground-state peak, while positive fields initially decrease it before recovery.

To provide a broader view, Fig. 6 plots the external field dependence of the split ground-state peak energy along the *x*, *y*, and *z* directions, assuming a 10 T internal field in the same direction. The *y*-direction internal field has the strongest impact, shifting the energy by ~ 1.5 meV, followed by the *x* (~ 1.0 meV) and *z* (~ 0.6 meV) directions. The slopes of the field dependencies follow the same trend, with the *y*-direction being the steepest (~ 0.2 meV/T) and the *z*-direction the weakest (~ 0.08 meV/T). Notably, applying an external magnetic field of ~ 7.8 T along the three principal axes approximately compensates for the peak shifts induced by the internal exchange field. Specifically, the CF excitations return to their original positions under this specific field magnitude, regardless of the field direction—a result that underscores the dominant influence of the external field on the energy landscape in $NdFeO_3$. However, it is important to emphasize that this 'compensating' external field should not be interpreted as a direct measurement of the internal exchange field strength. Rather, it reflects a complex rebalancing of the total magnetic environment acting on the $Nd^{3+}$ site. Interestingly, external and internal fields along the same direction exhibit opposite effects, warranting further investigation.

**Table 2.** Fitted CF parameters of $Nd^{3+}$ in $NdFeO_3$

| Param | $B_2^0$ | $B_2^2$ | $B_2^{-2}$ | $B_4^0$ | $B_4^2$ | $B_4^{-2}$ |
|---|---|---|---|---|---|---|
| Value | -3.3612e-01 | 2.1816e-01 | 5.7790e-01 | -2.9663e-03 | -4.40396e-03 | 2.1376e-02 |
| Param | $B_4^4$ | $B_4^{-4}$ | $B_6^0$ | $B_6^2$ | $B_6^{-2}$ | $B_6^4$ |
| Value | 1.9065e-02 | -1.2384e-02 | 3.4398e-04 | 6.5916e-04 | 1.2008e-03 | 3.3845e-04 |
| Param | $B_6^{-4}$ | $B_6^6$ | $B_6^{-6}$ | | | |
| Value | -7.4295e-04 | -6.4225e-04 | 7.5609e-04 | | | |

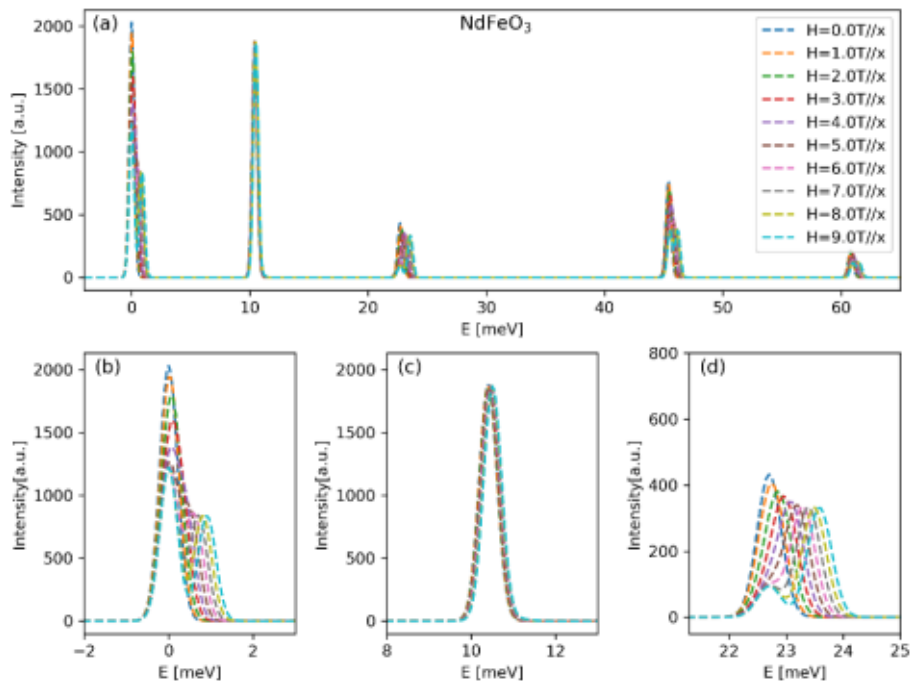

**Fig. 1** (a) CF excitation spectra of $Nd^{3+}$ in $NdFeO_3$ at different internal magnetic fields along the *x*-direction. (b), (c), and (d) show zoomed-in views of the first, second, and third CF excitation peaks of $Nd^{3+}$ from (a), respectively.

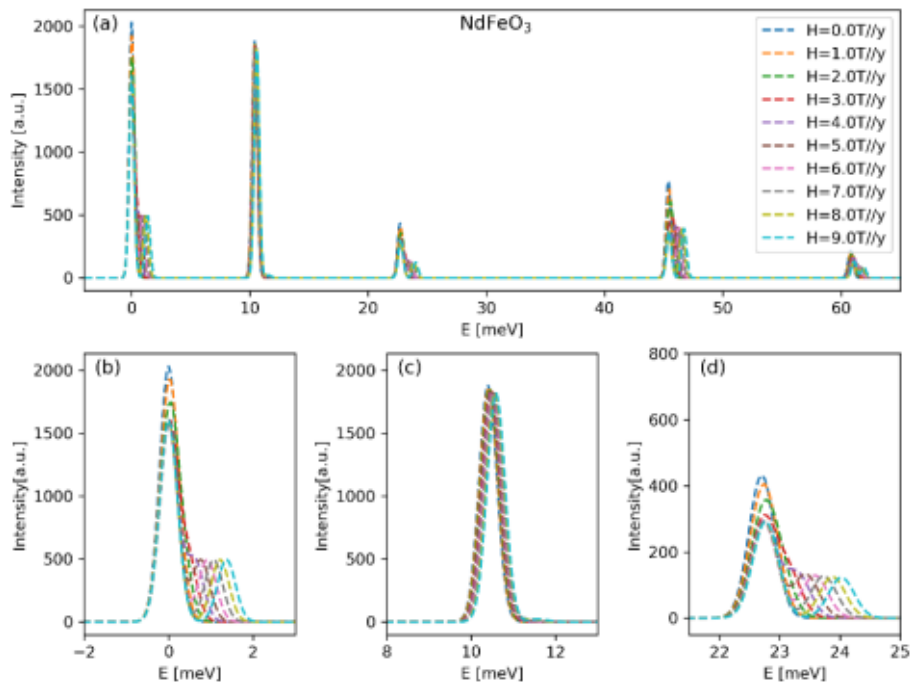

**Fig. 2** (a) CF excitation spectra of $Nd^{3+}$ in $NdFeO_3$ at different internal magnetic fields along the *y*-direction. (b), (c), and (d) show zoomed-in views of the first, second, and third CF excitation peaks of $Nd^{3+}$ from (a), respectively.

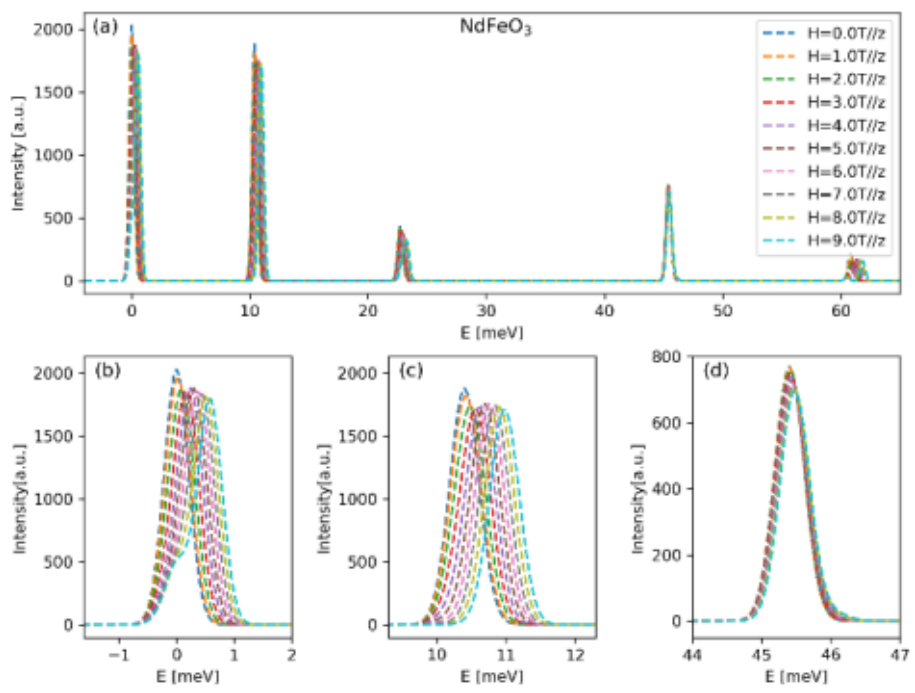

**Fig. 3** (a) CF excitation spectra of $Nd^{3+}$ in $NdFeO_3$ at different internal magnetic fields along the *z*-direction. (b), (c), and (d) show zoomed-in views of the first, second, and fourth CF excitation peaks of $Nd^{3+}$ from (a), respectively.

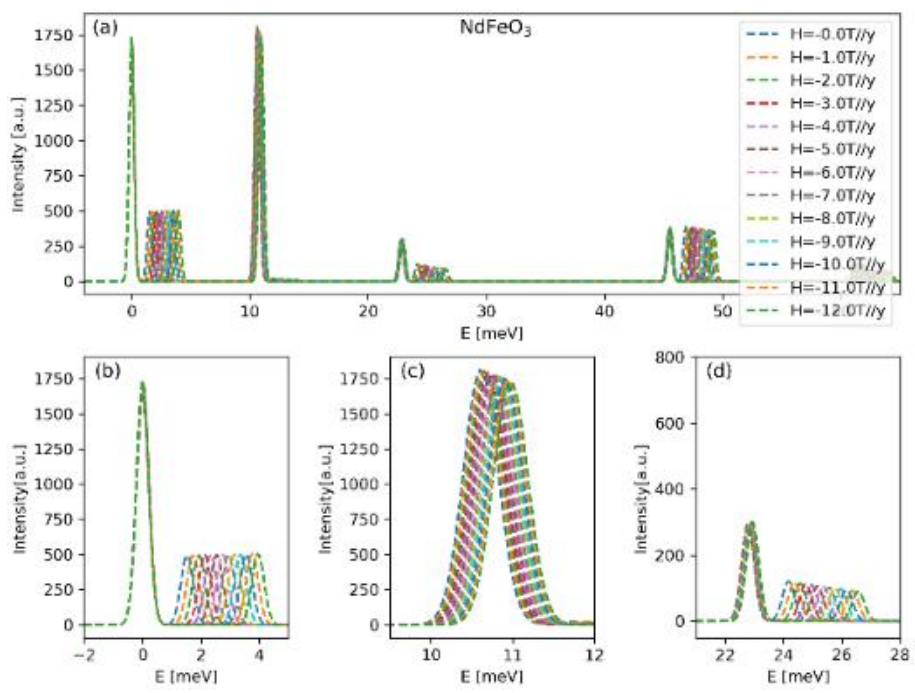

**Fig. 4** (a) CF excitation spectra of $Nd^{3+}$ in $NdFeO_3$ at different external fields along the *opposite* direction, assuming a 10 T internal field along the *y* direction. (b), (c), and (d) show zoomed-in views of first, second, and third CF peaks of $Nd^{3+}$ from (a), respectively.

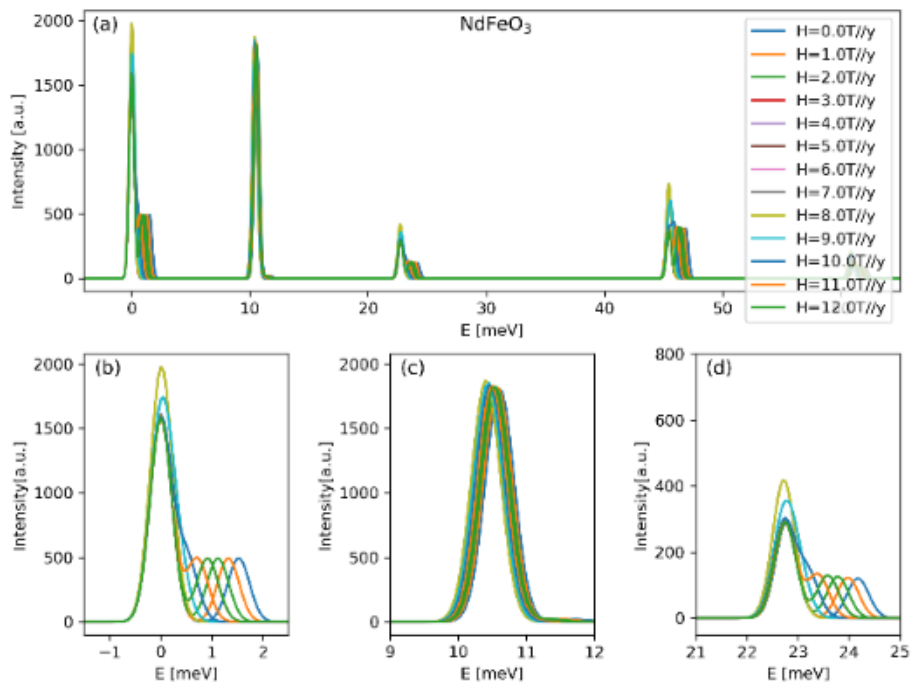

**Fig. 5** (a) CF excitation spectra of $Nd^{3+}$ in $NdFeO_3$ at different external fields along the *same* direction, assuming a 10 T internal field along the *y* direction. (b), (c), and (d) show zoomed-in views of first, second, and third CF peaks of $Nd^{3+}$ from (a), respectively.

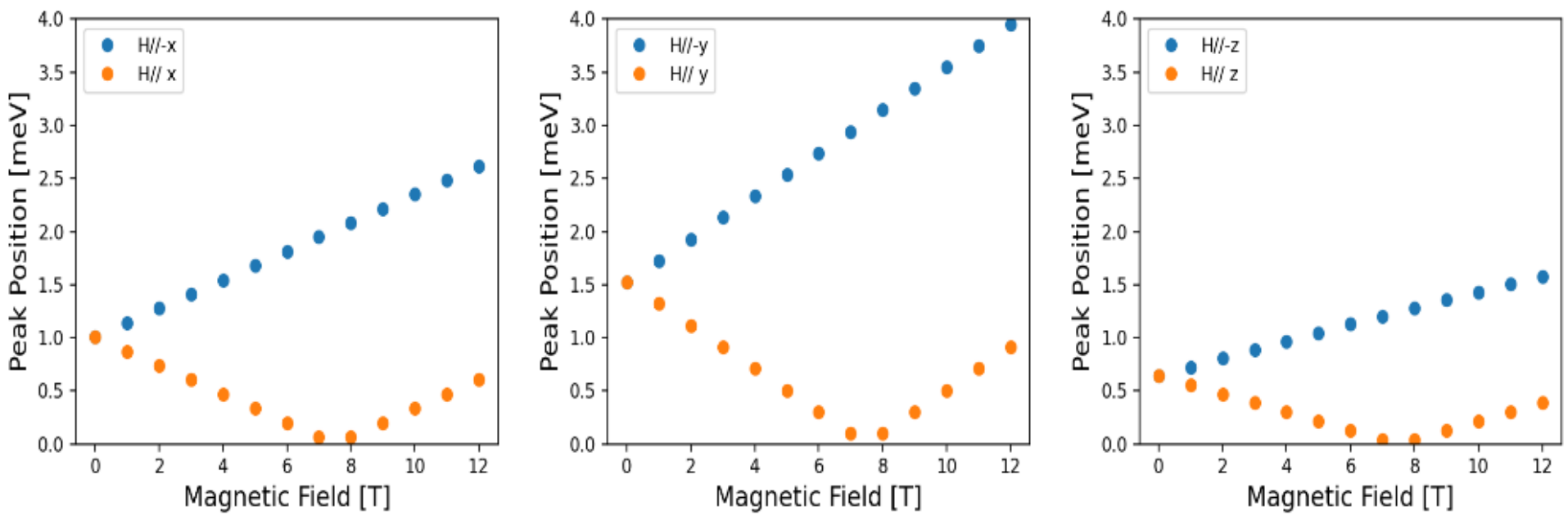

**Fig. 6** (a), (b), and (c) show the field dependence of the split ground-state CF excitation energy of $Nd^{3+}$ in $NdFeO_3$ under external fields applied in the same and opposite directions as the 10 T internal field along the *x*-, *y*-, and *z*-directions, respectively.

### 3.1.2 $ErFeO_3$

In our previous work, we have measured the CF excitations of $Er^{3+}$ from $ErFeO_3$. The observed excitation peaks are listed in Table 3.[11] With the initial parameters calculated from the point-charge model of $ErFeO_3$, we are able to fit the CF model to the observed peaks. The CF parameters generated from the model fitting are listed in Table 4.[11] With these parameters, we calculated the internal field and external field dependencies of the CF excitation peak positions. The obtained crystal excitation spectra with different internal fields can be found in our previous work.[3] These results clearly shows that the CF excitation peaks of $Er^{3+}$ are strongly dependent on the internal fields. Most of the peaks shift to higher energies while some peaks split as well. Here we also simulated the external field dependencies of the whole excitation spectra. The results are plotted in Figs. 7, 8, and 9. These three figures show that the external fields have stronger impacts on the excitation peaks than the internal magnetic fields do. When the external magnetic fields are applied along the *x* and *y* directions, all the CF excitation peaks split and significantly shift. However, when applying magnetic fields along the *z* direction, strong impacts such as peak splitting and energy shifting are observed for all excitation peaks. The impact on the ground state excitation is minimal while the changes to the second excitation peak are significant. The second peak not only splits but also shifts in a large energy scale in response to the external fields. Comparing all the effects of the fields along the three directions, the magnetic fields along the *y* direction have the strongest impact on the ground state excitation peak, causing the largest energy shift.

In our previous work, we also measured the field dependencies of the CF peak positions using inelastic neutron scattering technique, demonstrating some interesting Zeeman behaviours of the ground field excitation peak of $Er^{3+}$.[11] This peak firstly split into two peaks under the external fields. One of the split peaks moves to the higher energy side and becomes weaker and weaker and finally disappears, while the other peak shifts to the lower energy firstly and then move to the higher energies after approaching the lowest energy and gradually become stronger and stronger at high magnetic fields. The results demonstrate how the Zeeman effect impacts the CF peaks when the internal fields exist in $ErFeO_3$.

To confirm this observation, we simulated the $Er^{3+}$ CF spectra at different external fields along the *x*, *y,* and *z* directions by assuming the internal field is 10 T along the same directions, respectively. The magnetic fields along the *y* direction have the strongest impacts. Thus, we plot the results from the positive and negative *y* fields in Figs. 10 and 11. These two figures clearly exhibit that the negative fields firstly drive the excitation peak to the low energy side and then shift the peak to higher energy. However, the positive fields drive the excitation peak to the high energy monotonically. This indicates the internal and external fields have the similar impacts to the excitations. The negative and positive fields split both the ground state and the third excitation peaks. The split peak on the high energy side shifts significantly with the enhanced magnetic fields. The non-split first excitation peak only slightly shifts with the external field. The overall behaviour of the ground state excitation repeats the experimental observation of the Zeeman effect in this compound.[11] The field dependencies of the ground state excitation peak along the three directions are plotted in Fig. 12. Clearly, the external magnetic fields along the *y* direction have strongest impact on the excitation energy of the ground state. The excitation energy at zero external magnetic fields is at ~ 1.2 meV along the *y* direction, ~ 0.8 meV along the *x* direction, and ~ 0.2 meV along the *z* direction. At an external magnetic field of 9T, the peak positions are ~ 4.5, ~ 2.2, and ~ 0.7 meV under *y*, *x*, and *z*-directed fields, respectively, confirming that the internal field has the strongest effect along the *y* direction. Notably, the external field response becomes nonlinear above 5 T, consistent with Zeeman experiments on Sika. [11] The strong agreement between experimental and calculated results validates our CF model for $Er^{3+}$ in $ErFeO_3$. Additionally, the minimum excitation energy occurs around 3 T in all directions, suggesting that an external field of ~3T appears to cancel the effect of the internal 10 T field. This indicates that the internal field effect in $ErFeO_3$ is significantly weaker than in $NdFeO_3$.

Table 3. Calculated CF peak positions and intensities of $Er^{3+}$ Kramers doublets in $ErFeO_3$

| | | | | | |
|---|---|---|---|---|---|
| Energy [meV] | 0.0 | 5.79 | 13.43 | 20.47 | 25.64 |
| Intensity [a.u.] | 758.94 | 1817.53 | 1209.69 | 438.49 | 172.24 |

Table 4. Fitted CF parameters of $Er^{3+}$ in $ErFeO_3$

| | | | | | | |
|---|---|---|---|---|---|---|
| Param | $B_2^0$ | $B_2^2$ | $B_2^{-2}$ | $B_4^0$ | $B_4^2$ | $B_4^{-2}$ |
| Value | -0.3313 | -0.526 | 8.275×10-7 | 3.21×10-4 | 7.318 ×10-3 | -1.876×10-4 |
| Param | $B_4^4$ | $B_4^{-4}$ | $B_6^0$ | $B_6^2$ | $B_6^{-2}$ | $B_6^4$ |
| Value | 4.15×10-3 | -1.821×10-4 | 2.87×10-6 | 1.074×10-4 | 9.563×10-06 | -1.501×10-4 |
| Param | $B_6^{-4}$ | $B_6^6$ | $B_6^{-6}$ | | | |
| Value | -9.931×10-6 | 1.498×10-4 | 4.394×10-6 | | | |

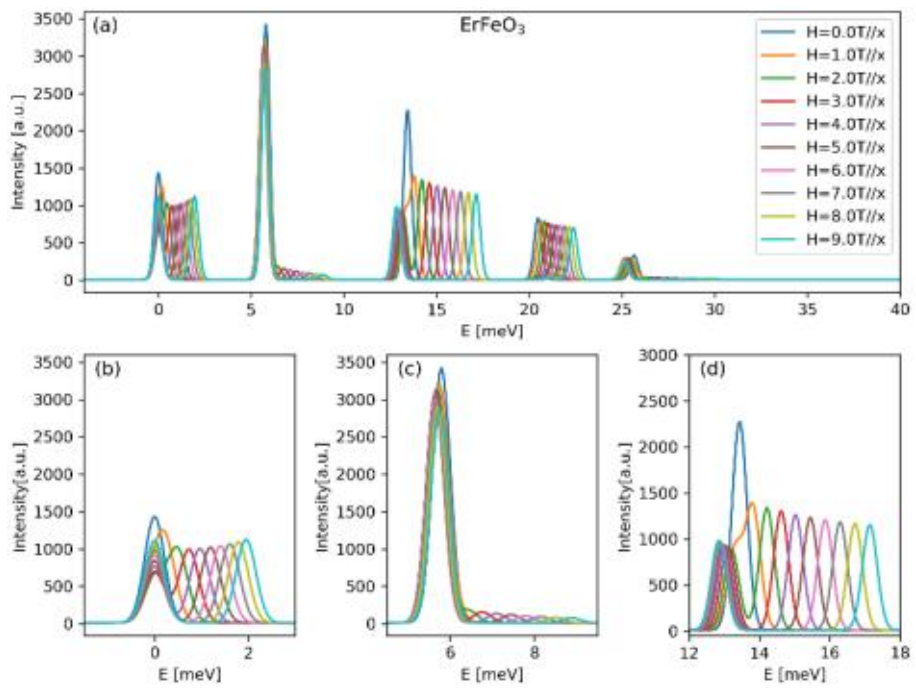

**Fig. 7** CF excitation spectra of $Er^{3+}$ in $ErFeO_3$ at different internal magnetic fields along the *x*-direction. (b), (c), and (d) show zoomed-in views of the first, second, and third CF excitation peaks of $Er^{3+}$ from (a), respectively.

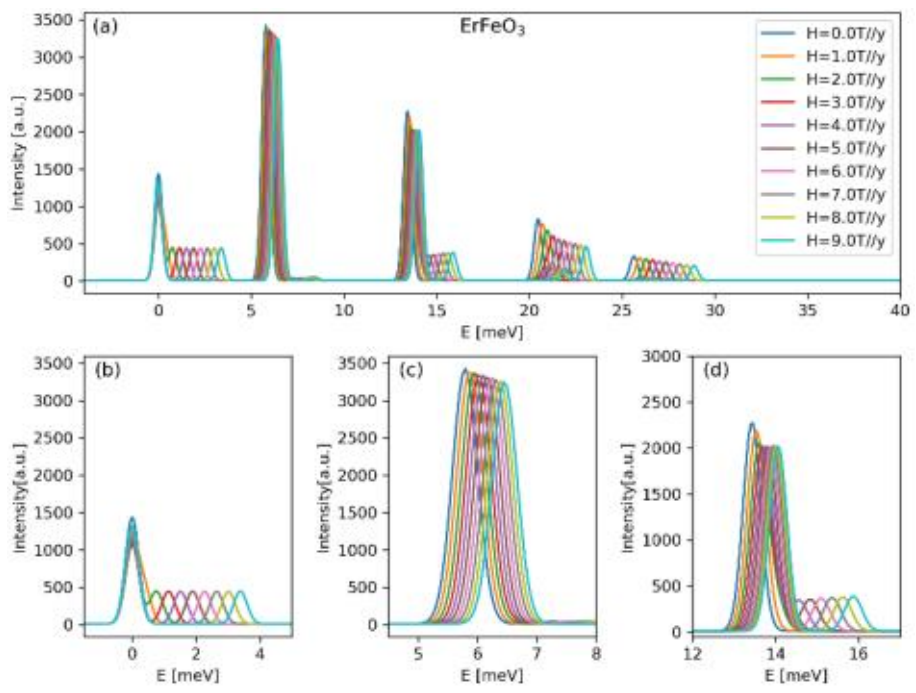

**Fig. 8** CF excitation spectra of $Er^{3+}$ in $ErFeO_3$ at different internal magnetic fields along the *y*-direction. (b), (c), and (d) show zoomed-in views of the first, second, and third CF excitation peaks of $Er^{3+}$ from (a), respectively.

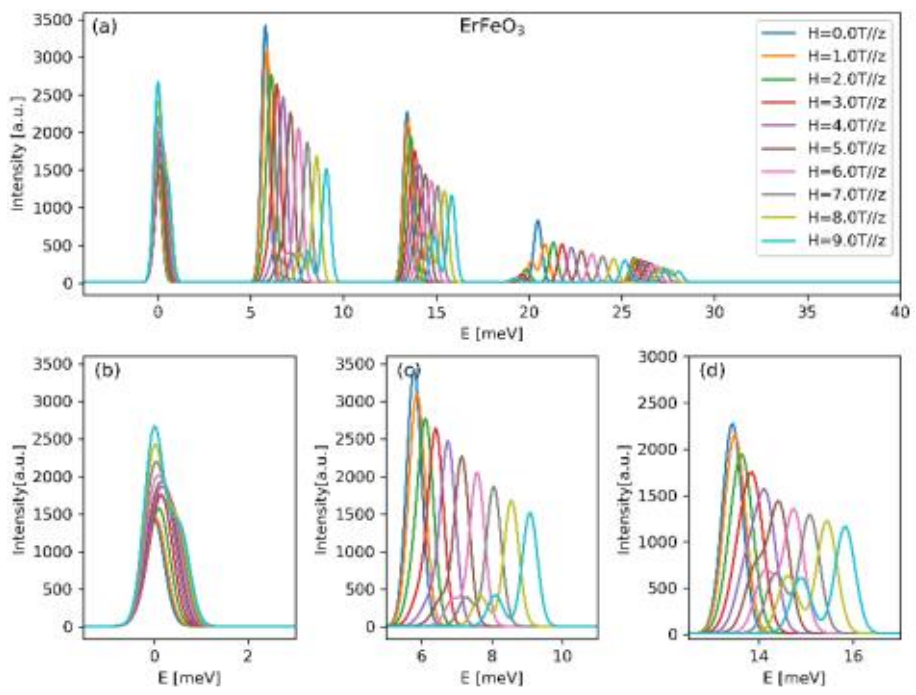

**Fig. 9** (a) CF excitation spectra of $Er^{3+}$ in $ErFeO_3$ at different internal magnetic fields along the *z*-direction. (b), (c), and (d) show zoomed-in views of the first, second, and third CF excitation peaks of $Er^{3+}$ from (a), respectively.

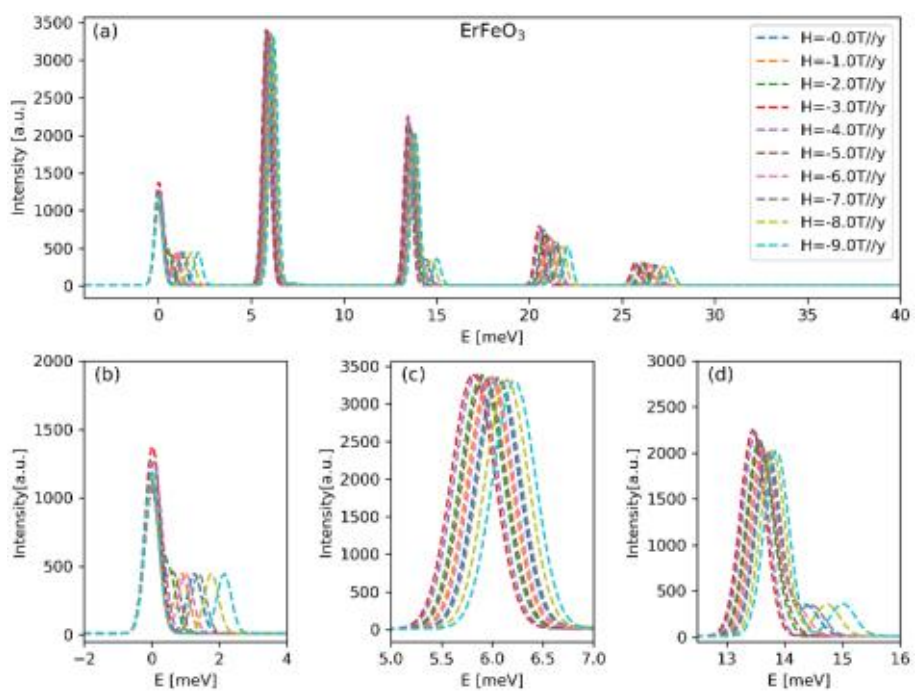

**Fig. 10** (a) CF excitation spectra of $Er^{3+}$ in $ErFeO_3$ at different external fields along the *opposite* direction, assuming a 10 T internal field along the *y* direction. (b), (c), and (d) show zoomed-in views of first, second, and third CF peaks of $Er^{3+}$ from (a), respectively.

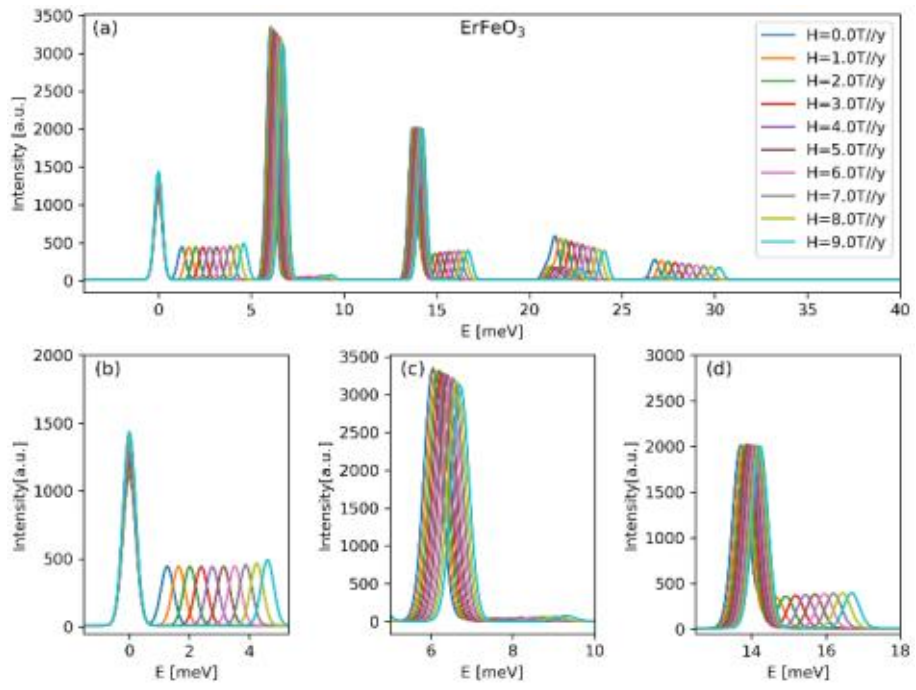

**Fig. 11** (a) CF excitation spectra of $Er^{3+}$ in $ErFeO_3$ at different external fields along the *same* direction, assuming a 10 T internal field along the *y* direction. (b), (c), and (d) show zoomed-in views of first, second, and third CF peaks of $Er^{3+}$ from (a), respectively.

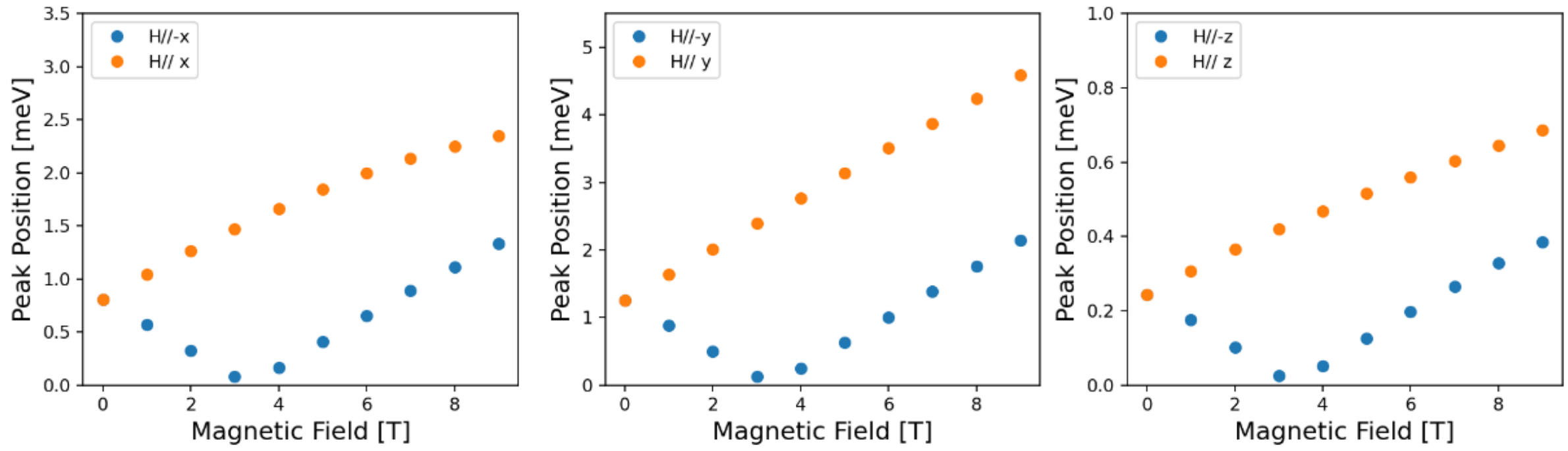

**Fig. 12** (a), (b), and (c) show the field dependence of the split ground-state CF excitation energy of $Er^{3+}$ in $ErFeO_3$ under external fields applied in the same and opposite directions as the 10 T internal field along the *x*-, *y*-, and *z*-directions, respectively.

### 3.1.3 $YbFeO_3$

$Yb^{3+}$ in $YbFeO_3$ has the $4f^{13}$ electronic configuration with $S = 1/2$ and $J = 7/2$. The ground state of $Yb^{3+}$ denotes as $^2F_{7/2}$. According to the Kramers degeneracy theorem, in a magnetic system with a half-integer total spin, each energy eigenstate of this system is accompanied by another eigenstate with the same energy due to the time-reversal symmetry. Namely, the CF excitation levels of $Yb^{3+}$ consist of a series of degenerated doublets. With the $C_s$ local symmetry in $YbFeO_3$, the ground state of $Yb^{3+}$ splits into $(J + ½) = 4$ doublet states due to the CF effect of the surrounding ligands. According to the previous report by S. E. Nikitin et al.,[8] $Yb^{3+}$ in $YbFeO_3$ demonstrates a strong CF excitation peak at ~20 meV. At low temperature, a low-lying CF excitation near 0.5 meV was observed at 10K while this peak evolved to ~ 1 meV with a weak dispersion along the *c* axis at 2 K. The external magnetic fields drove this excitation peak to higher energies. A similar CF excitation was observed at ~30 meV in another similar compound $YbAlO_3$ by L. S. Wu et al.[29] They calculated the CF excitation energy levels using a point-charge model, which resulted in the first excitation energy at 29.7 meV, consistent with the experimental value. As the $Yb^{3+}$ moments form the long-range magnetic order at 2K, a gapped dispersive excitation was observed at a low energy ~ 0.5 meV. The dispersion could be attributed to the exchange interaction between the ordered $Yb^{3+}$ magnetic moments. However, the energy gap could be induced by the shift of the ground-state CF excitation due to internal magnetic fields. Both the $Fe^{3+}$ and $Yb^{3+}$ sublattices may play roles in creating such internal magnetic fields, just as the $Er^{3+}$ low-lying excitation in $ErFeO_3$.

From Nikitin's report,[30] we could obtain the excitation energies and the relative intensities of the crystal levels of $YbFeO_3$, as shown in Table 5. Fitting the CF excitation model to the excitation energies and intensities generates the CF parameters shown in Table 6. Using these parameters, the internal magnetic field dependencies of the excitation spectra of $YbFeO_3$ are calculated along the *x*, *y*, and *z* directions and shown in Figs. 13, 14, and 15, respectively.

Fig. 13 shows that an internal field along the *x* direction splits and shifts the ground-state excitation slightly above zero energy, with minor shifts in the other two peaks. Similarly, in Fig. 14, an internal field along the *y* direction causes comparable shifts (less than 2 meV) in the ground-state and first excitation peaks (~20 meV), but the third peak (~53 meV) undergoes a surprisingly large splitting. In contrast, Fig. 15 reveals that internal fields along the *z* direction induce the largest energy shifts across all three peaks, though no splitting occurs for the third excitation peak. These results indicate that an internal field in any direction shifts the ground-state excitation away from zero to higher energy. As reported, a low-energy $Yb^{3+}$ excitation was observed in the magnetically ordered phase of $YbAlO_3$, similar to the one observed in $YbFeO_3$, which indicates that both the $Fe^{3+}$ and $Yb^{3+}$ sublattices induce internal magnetic fields when they order.

Similar calculated results for the external magnetic field are presented in Figs. S4, S5, and S6 in the Supplemental Material. In general, the external magnetic fields have similar but more pronounced impacts on the CF excitations of $Yb^{3+}$ in $YbFeO_3$.

The Zeeman effect is simulated for $YbFeO_3$ with a 10 T internal field along the *z* direction. The CF spectra in fields along the -*z* and *z* directions are shown in the Figs. 16 and 17, respectively. We see that the opposite external fields cancel some of the internal field effect and finally drive the peak to the higher energy. While the external magnetic fields in the same direction monotonically drive the field to higher energies.

The Zeeman effect along the *x*, and *y* directions are also simulated. The Zeeman effects along the *x*, *y*, and *z* directions are summarized in Fig. 18 (a), (b), and (d), respectively. No matter in which direction the magnetic external fields are, the minimal energies of the CF excitation peaks are all at ~ 2.5 T. Namely, an external field at this level can cancel the internal fields of 10 T in the opposite directions, which is like the case of $Er^{3+}$ in $ErFeO_3$ but much smaller than the field needed for $Nd^{3+}$ in $NdFeO_3$.

**Table 5.** CF excitation levels of $YbFeO_3$ [30]

| Energy [meV] | 0 | 20 | 53.3 | 64.7 |
|---|---|---|---|---|
| Intensity [a.u.] | 360 | 308 | 28 | 8 |

**Table 6.** CF parameters of $Yb^{3+}$ by fitting to the peaks and intensities in Table 5

| Param | $B_2^0$ | $B_2^2$ | $B_2^{-2}$ | $B_4^0$ | $B_4^2$ | $B_4^{-2}$ |
|---|---|---|---|---|---|---|
| Value | -1.4989e+00 | -2.5173e-01 | 1.3098e+00 | 9.3083e-03 | -7.7327e-04 | 5.9473e-02 |
| Param | $B_4^4$ | $B_4^{-4}$ | $B_6^0$ | $B_6^2$ | $B_6^{-2}$ | $B_6^4$ |
| Value | 6.4702e-02 | -1.4093e-02 | 1.9878e-04 | 2.9255e-04 | -2.0802e-03 | -2.2709e-03 |
| Param | $B_6^{-4}$ | $B_6^{6}$ | $B_6^{-6}$ | | | |
| Value | 9.9630e-04 | -2.1636e-03 | -4.72E-03 | | | |

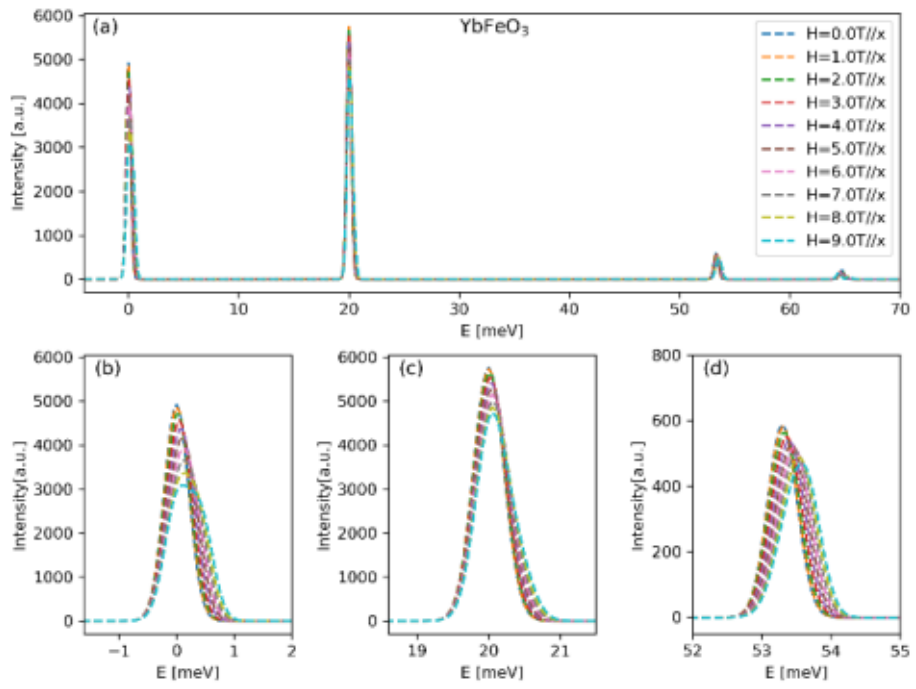

**Fig. 13** CF excitation spectra of $Yb^{3+}$ in $YbFeO_3$ at different internal magnetic fields along the *x*-direction. (b), (c), and (d) show zoomed-in views of the first, second, and third CF excitation peaks of $Yb^{3+}$ from (a), respectively.

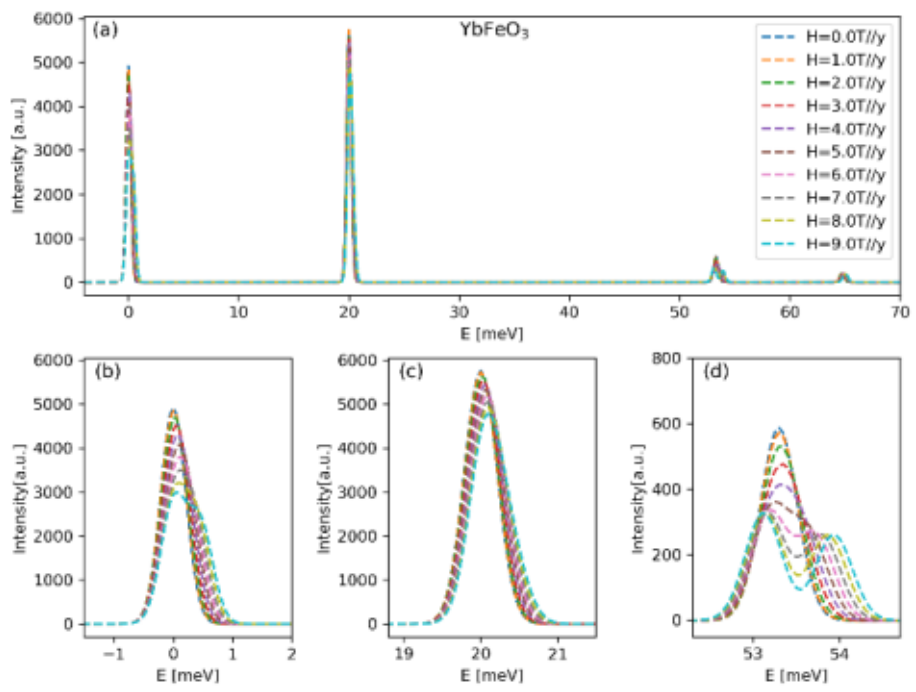

**Fig. 14** CF excitation spectra of $Yb^{3+}$ in $YbFeO_3$ at different internal magnetic fields along the *y*-direction. (b), (c), and (d) show zoomed-in views of the first, second, and third CF excitation peaks of $Yb^{3+}$ from (a), respectively.

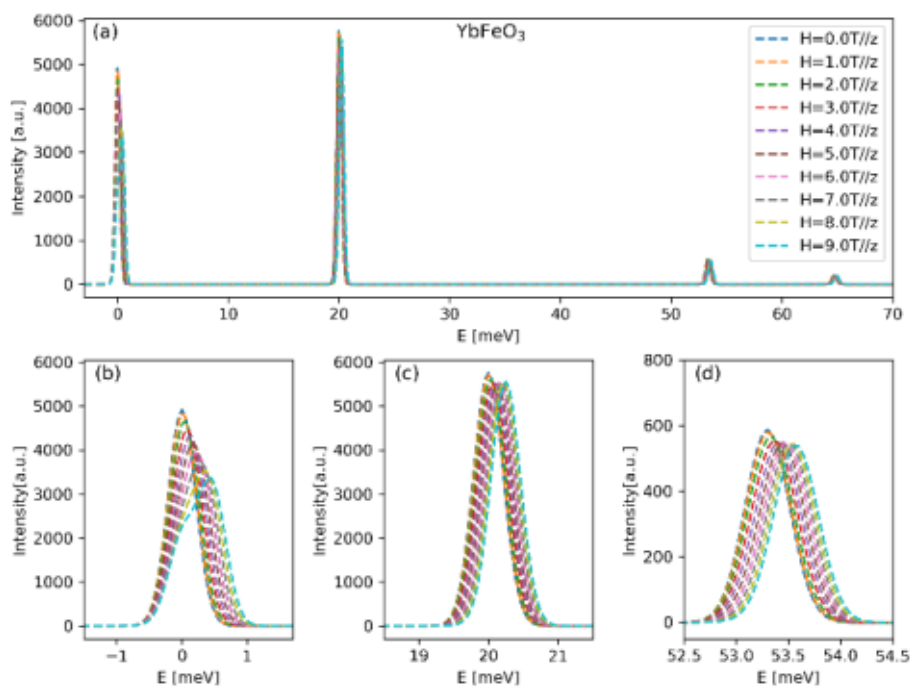

**Fig. 15** CF excitation spectra of $Yb^{3+}$ in $YbFeO_3$ at different internal magnetic fields along the *z*-direction. (b), (c), and (d) show zoomed-in views of the first, second, and third CF excitation peaks of $Yb^{3+}$ from (a), respectively.

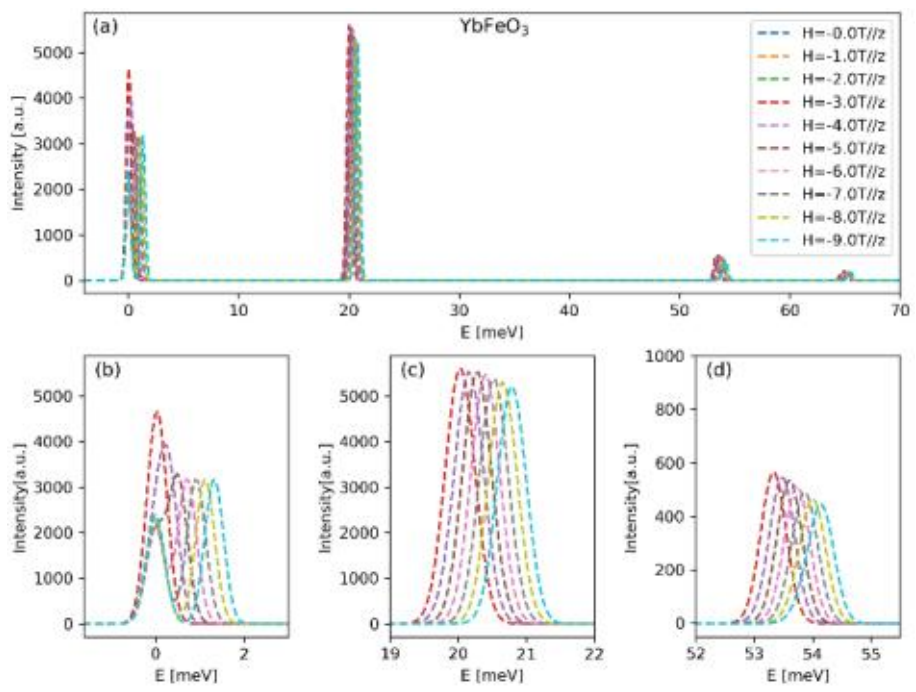

**Fig. 16** (a) CF excitation spectra of $Yb^{3+}$ in $YbFeO_3$ at different external fields along the *opposite* direction, assuming a 10 T internal field along the *z* direction. (b), (c), and (d) show zoomed-in views of first, second, and third CF peaks of $Yb^{3+}$ from (a), respectively.

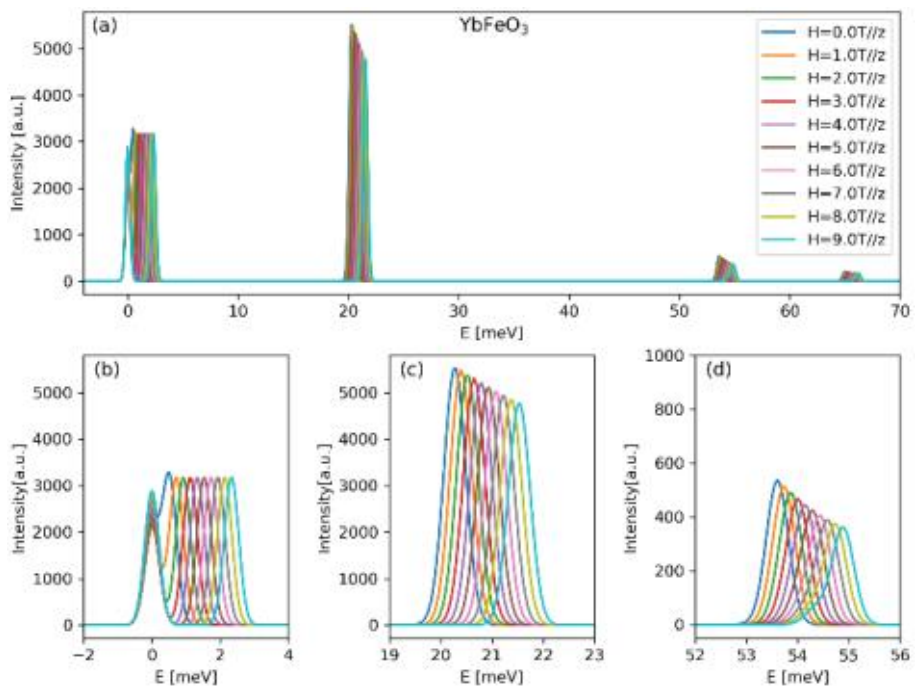

**Fig. 17** (a) CF excitation spectra of $Yb^{3+}$ in $YbFeO_3$ at different external fields along the *same* direction, assuming a 10 T internal field along the *z* direction. (b), (c), and (d) show zoomed-in views of first, second, and third CF peaks of $Yb^{3+}$ from (a), respectively.

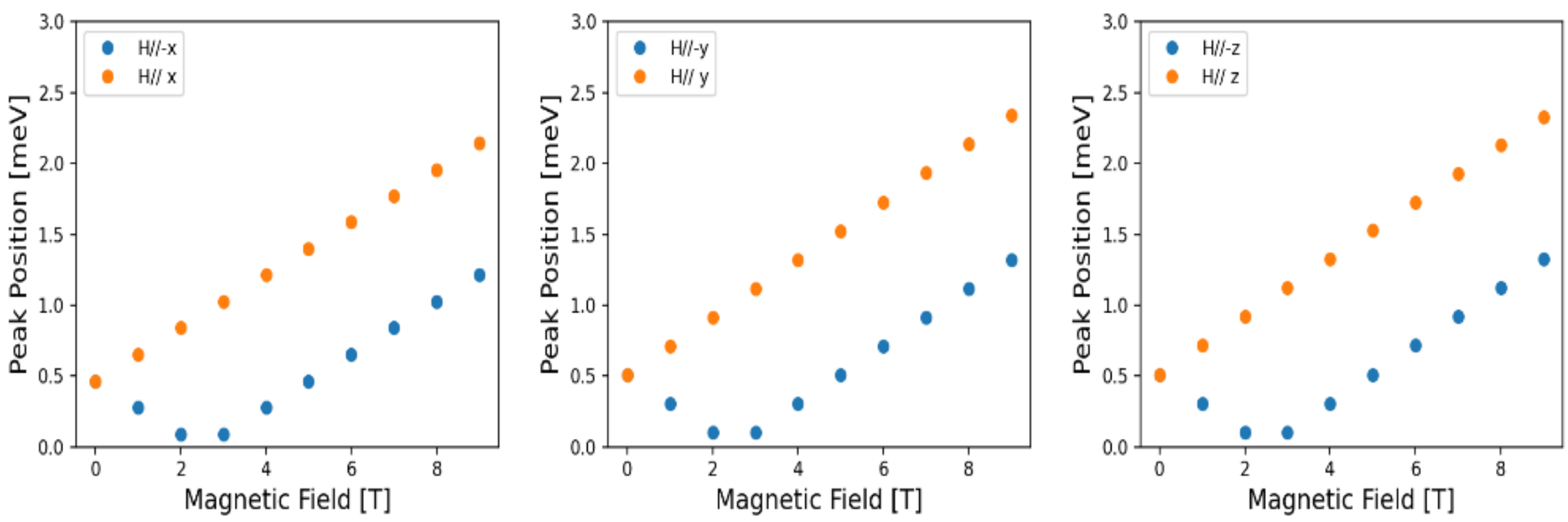

**Fig. 18** (a), (b), and (c) show the field dependence of the split ground-state CF excitation energy of $Yb^{3+}$ in $YbFeO_3$ under external fields applied in the same and opposite directions as the 10 T internal field along the *x*-, *y*-, and *z*-directions, respectively.

### 3.2 *Non-Kramers ions $Pr^{3+}$ and $Ho^{3+}$ in $REFeO_3$*

#### 3.2.1 $PrFeO_3$

$Pr^{3+}$ in $PrFeO_3$ has the $4f^2$ electronic configuration with $S = 1$ and $J = 4$. The ground state of $Pr^{3+}$ denotes as $^3H_4$. Since $Pr^{3+}$ has an integer total spin *S*, it is a non-Kramers' ion. The CF excitation of $Pr^{3+}$ does not follow the Kramers degeneracy theorem. In the case of non-Kramers' rare earth ions, the CF effect completely removes the degeneracy of the energy sublevels, resulting in $(2J + 1)$ singlets for the ground state multiplet. This means that the ground state $^3H_4$ of $Pr^{3+}$ splits into 9 singlet energy levels in the $C_s$ local symmetry of $PrFeO_3$. According to the irreducible representation analysis, these 9 singlets can be classified into two different irreps: A' (symmetric) and A" (asymmetric). The CF excitations of $PrFeO_3$ and $PrGaO_3$ were experimentally studied earliest by K. Feldmann et al.[31] using inelastic neutron scattering technique. They observed excitations at 2.0, 14.7, 23.2, 36, and 58 meV from $PrFeO_3$, see Table 7. The excitation peaks from $PrGaO_3$ were in the similar range with some energy shifts.

We developed a CF model to fit the CF peaks of $Pr^{3+}$ observed in $PrFeO_3$. The initial CF parameters were initialized using the point-charge model and refined them to match the observed peaks. According to our analysis, the peak at ~36 meV does not correspond to a ground state CF excitation but rather a transition between excited states (~23 meV to ~58 meV), confirmed by its absence at 10K and appearance at 80K. The model successfully fits the remaining peaks, generating the optimized CF parameters in Table 8. Table 9 presents the eigenstates. peak positions, and intensities of the CF excitations. The results show seven well-separated singlets and a pseudo-doublet at ~14.7 meV. The excitation at ~0.87 meV is nearly silent while the peak at ~2 meV is the strongest.

Using the fitted CF parameters, internal field effects are simulated for $PrFeO_3$ by applying magnetic fields along the $x$, $y$, and $z$ directions, with results shown in Figs. 19, 20, and 21. When the internal field is along the $x$ direction, most peaks exhibit significant shifts and intensity changes. The most notable effect occurs at zero energy transfer, where no peak is present at zero field. As the internal field increases, a new excitation peak emerges at zero energy transfer and strengthens, along with the ~0.87 meV peak, which is nearly invisible at zero field but gains intensity with increasing field. Energy shifts are observed for peaks at 2.0 meV, 14.7 meV, 23 meV, and 58 meV. The only observed splitting occurs at ~14.7 meV, consistent with its pseudo-doublet nature. We attribute this splitting to the lifting of accidental degeneracy, as confirmed by the eigenstates in Table 9.

The internal fields along the $y$ direction have much less impacts to the excitation peaks in comparison to the fields along $x$. Even though Fig. 20 clearly presents the enhanced peak intensity at zero energy transfer, its total intensity is still very weak, not comparable to the observed ones in the $x$ fields at all. The peak at ~0.87meV is hardly to be observed. The rest excitation peaks show almost no changes. As shown in Fig. 21, the internal magnetic fields along the $z$ direction almost have no impacts to all the excitation peaks at all, even to the ground state excitation.

We also simulated external magnetic field effects using the same model. The simulated results are shown in Figs. S7, S8, and S9. In general, the external magnetic fields demonstrate similar, but stronger, effects as the internal fields.

It is interesting to investigate the external magnetic field effects along three different directions for the existence of an internal field of 10 T in the same direction. We did simulate this scenario for all the three directions. We simulated the external magnetic field effect by applying a magnetic external field alone the $x$ direction and keeping an internal magnetic field of 10 T along the same direction. The magnetic fields are applied in the negative and positive directions relative to the internal magnetic field. The simulated results along the -$x$ and $x$ directions are plotted in Figs. 22 and 23, respectively.

We found the negative internal magnetic fields drive the excitation energy to an even higher level while the positive external magnetic fields firstly cancel the internal magnetic field effect and completely overcome them and then drive the excitation peaks to higher energies (see Fig. 24). To our surprise, such a behaviour is completely opposite to what we observed in $Er^{3+}$ and $Yb^{3+}$ above, but like the case of $Nd^{3+}$. However, the external magnetic fields along the $y$ direction exhibit normal effects. The fields along the $z$ direction still do not exhibit any impact to the peak positions and intensities.

Comparing to the Kramers' ions like $Nd^{3+}$, $Er^{3+}$, and $Yb^{3+}$ discussed above, we observed several different features in our simulated spectra for $Pr^{3+}$. The first feature is that no ground state excitation shows up at zero energy transfer under zero magnetic field. While the ground state excitations were observed in the cases of all the three Kramers' ions. Applying internal or external magnetic fields along *x* and *y* can induce the ground-state excitation at the zero-energy transfer. This can be explained by the non-Kramers' nature of $Pr^{3+}$. For a non-Kramers' ion like $Pr^{3+}$, it has a non-degenerate singlet as the ground state, which means that the excitation can only take place between the ground state to another excited state. However, in a Kramers ion, the ground state consists of a degenerated doublet, which makes the excitation possible from one degenerated state of the doublet to the other one if the transition matrix allows. According to the eigenstates of the ground state of $Nd^{3+}$, $Er^{3+}$, and $Yb^{3+}$, the two degenerated states of the ground state doublet can transit from one to another according to the selection rule. Thus, their ground state excitations at zero energy transfer can be observed in inelastic neutron scattering experiments. Contrarily, a non-Kramers' ion like $Pr^{3+}$, there is no change to transition from its singlet ground state to itself, resulting no excitation at zero energy transfer.

An intriguing feature of the CF excitation spectrum is the relatively weak ("almost-silent") peak at approximately 0.87 meV. Typically, excitations closer to the ground state exhibit greater intensity; however, simulations reveal that the peak at approximately 2 meV is significantly stronger than the one at 0.87 meV, despite its higher energy. Notably, previous experimental data did not detect this low-energy peak. This behaviour can be understood through selection rules governing magnetic dipole transitions. The ground state's dominant eigenvector component is $m_j = -4$, while the excited state at ~0.87 meV is primarily $m_j = -2$. The transition between these states ($\Delta m_j = 2$) is forbidden under the magnetic dipole selection rule ($\Delta m_j = 0, \pm1$), resulting in weak intensity. This faint signal likely arises from minor contributions of other mj components (e.g., $\Delta m_j = 1$ transitions), which are weakly allowed. In contrast, the excited state at ~2 meV shares the dominant $m_j = -4$ component with the ground state ($\Delta m_j = 0$), rendering this transition fully allowed and thus much more intense. This selection rule dependence explains the observed intensity disparity and suggests the simulation captures subtle effects potentially missed in earlier experiments.

The application of internal or external magnetic fields increases the intensities of the ground state excitation and the excitation at 0.87 meV, as exhibited in Figs. 19 ~ 21 and S7 ~ S9 of the Supplemental Material. Magnetic fields break the original symmetry of the system, disrupting selection rules tied to time-reversal or spatial symmetry and enabling transitions previously forbidden. As internal field strength increases, the magnetic moments become strongly polarized, enhancing transition dipole moments and thus intensifying the CF excitation peaks. For non-Kramers ions, a singlet ground state—potentially forming a pseudo-doublet due to accidental degeneracy—undergoes Zeeman splitting under magnetic fields, resulting in a set of closely spaced energy levels rather than a single state. Such splitting permits low-energy transitions within the ground state manifold, manifesting as an enhanced excitation near zero-energy transfer. Similarly, the transition from the ground state to the excited state at 0.87 meV becomes nonprohibited due to symmetry breaking, which relaxes selection rules and boosts its intensity. These effects underscore the role of magnetic fields in altering both state mixing and transition probabilities in $REFeO_3$ orthoferrites.

**Table 7.** CF excitation levels reported in $PrFeO_3$ with estimated intensities[32]

| Energy [meV] | 2.0 | 14.7 | 23.2 | 58 |
|---|---|---|---|---|
| Intensity [a.u.] | 50 | 30 | 20 | 12 |

**Table 8.** Fitted CF parameters of $Pr^{3+}$ in $PrFeO_3$

| Param | $B_2^0$ | $B_2^2$ | $B_2^{-2}$ | $B_4^0$ | $B_4^2$ | $B_4^{-2}$ |
|---|---|---|---|---|---|---|
| Value | 0.96466 | -0.30833 | 0.31463 | 3.5216e-03 | 4.7419e-2 | -2.6824e-2 |
| Param | $B_4^4$ | $B_4^{-4}$ | $B_6^0$ | $B_6^2$ | $B_6^{-2}$ | $B_6^4$ |
| Value | -2.4684e-02 | 1.8295e-02 | -7.1063e-05 | 2.2991e-03 | -1.8198e-03 | -1.6295e-03 |
| Param | $B_6^{-4}$ | $B_6^6$ | $B_6^{-6}$ | | | |
| Value | 4.5288e-03 | 2.0872e-03 | -3.0291e-03 | | | |

**Table 9.** Calculated peak positions and intensities of $PrFeO_3$

| | | | | | | | | | |
|---|---|---|---|---|---|---|---|---|---|
| Energy [meV] | 0 | 0.8750 | 2.008 | 14.699 | 14.715 | 23.204 | 42.759 | 58.005 | 78.115 |
| Intensity [a.u.] | 0 | 2.999 | 271.80 | 153.342 | 10.973 | 108.44 | 1.28 | 66.23 | 2.198 |

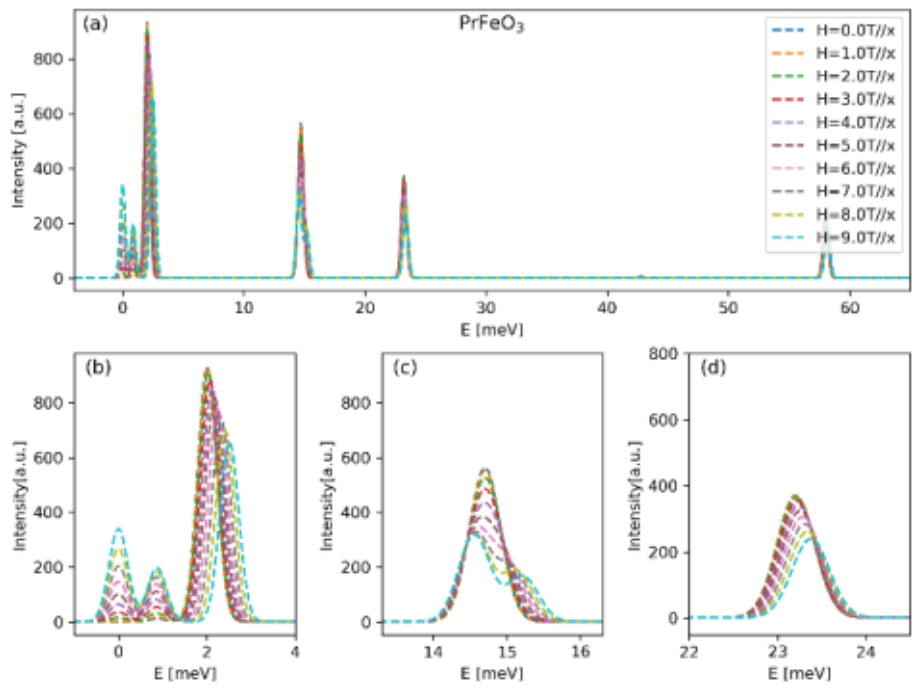

**Fig. 19** (a) CF excitation spectra of $Pr^{3+}$ in $PrFeO_3$ at different internal magnetic fields along the *x*-direction. (b), (c), and (d) show zoomed-in views of the first, second, and third CF excitation peaks of $Pr^{3+}$ from (a), respectively.

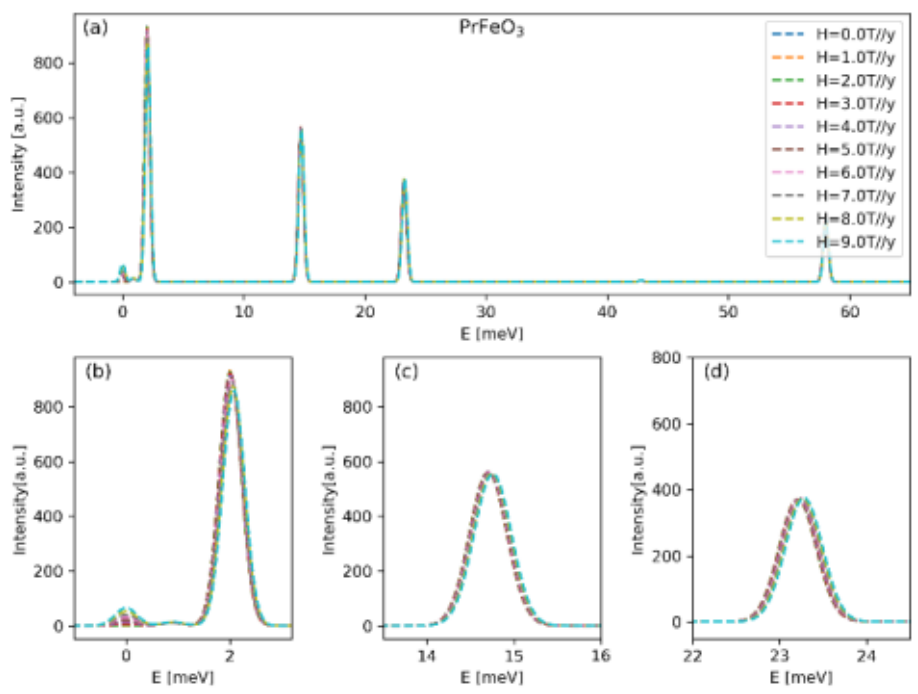

**Fig. 20** (a) CF excitation spectra of $Pr^{3+}$ in $PrFeO_3$ at different internal magnetic fields along the *y*-direction. (b), (c), and (d) show zoomed-in views of the first, second, and third CF excitation peaks of $Pr^{3+}$ from (a), respectively.

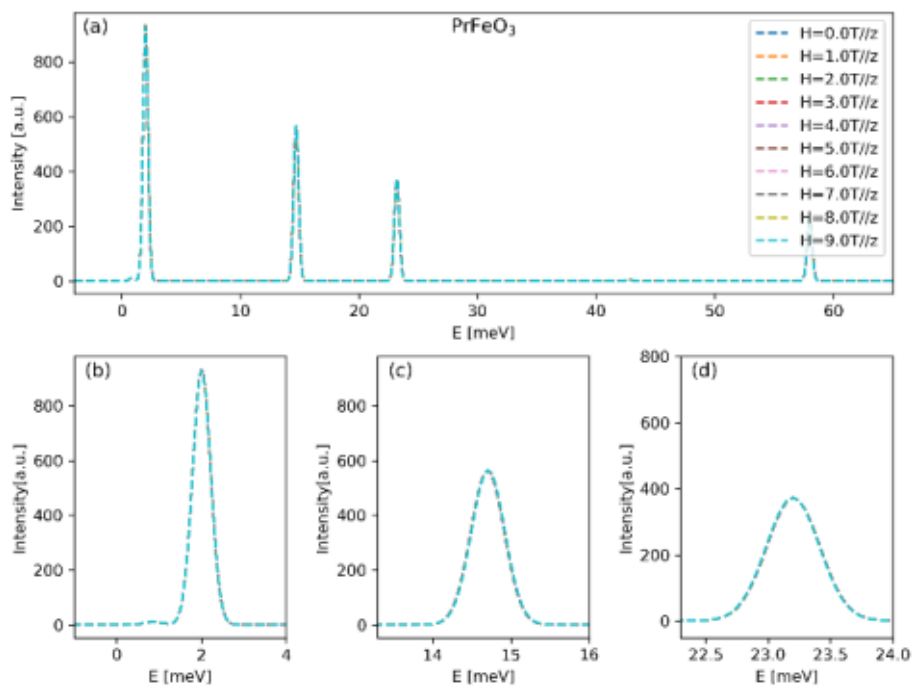

**Fig. 21** (a) CF excitation spectra of $Pr^{3+}$ in $PrFeO_3$ at different internal magnetic fields along the *z*-direction. (b), (c), and (d) show zoomed-in views of the first, second, and third CF excitation peaks of $Pr^{3+}$ from (a), respectively.

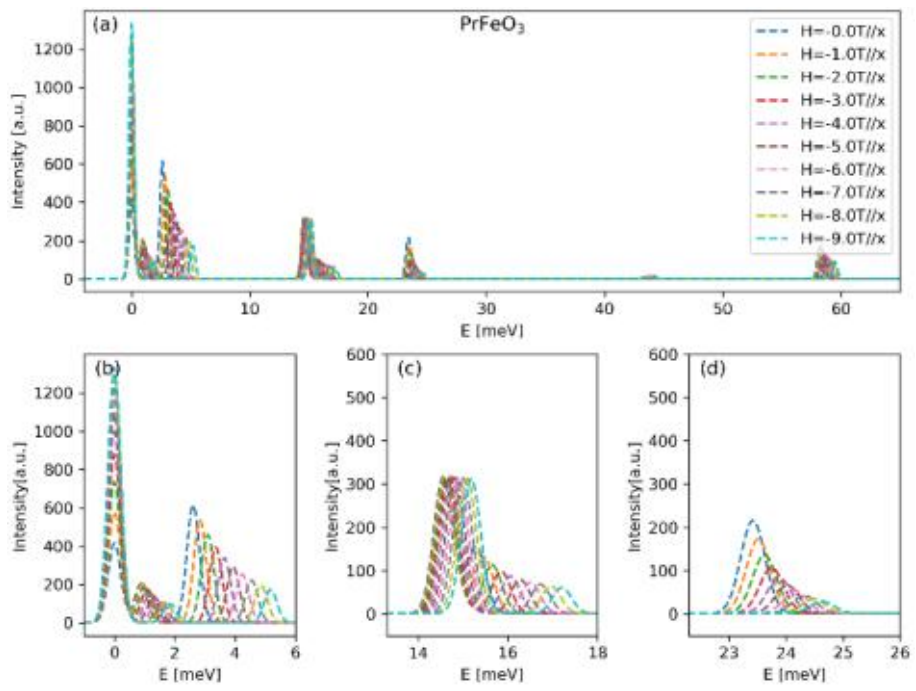

**Fig. 22** (a) CF excitation spectra of $Pr^{3+}$ in $PrFeO_3$ at different external fields along the *opposite* direction, assuming a 10 T internal field along the *x* direction. (b), (c), and (d) show zoomed-in views of first, second, and third CF peaks of $Pr^{3+}$ from (a), respectively.

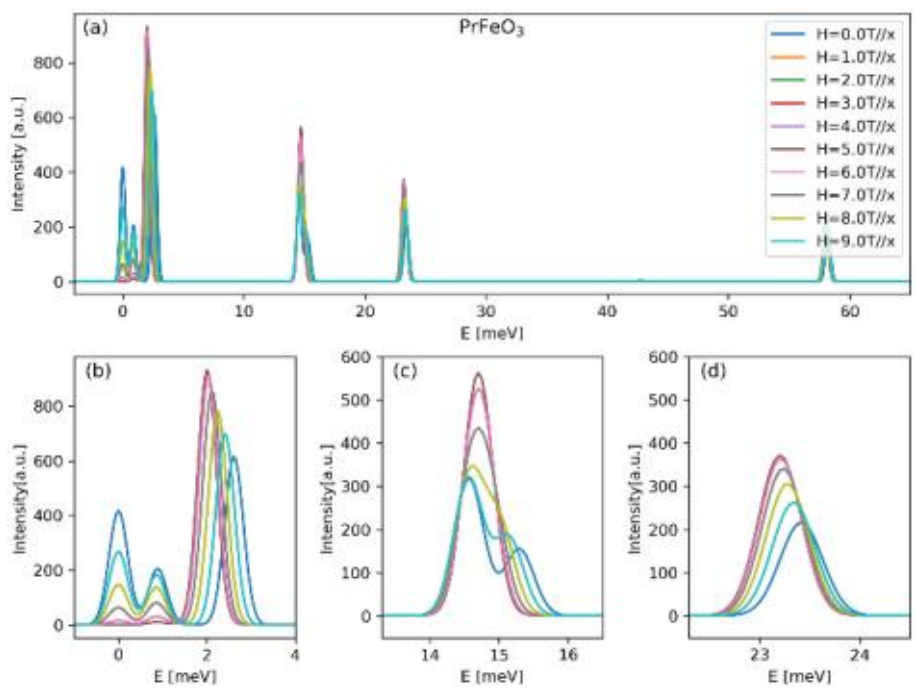

**Fig. 23** (a) CF excitation spectra of $Pr^{3+}$ in $PrFeO_3$ at different external fields along the *same* direction, assuming a 10 T internal field along the *x* direction. (b), (c), and (d) show zoomed-in views of first, second, and third CF peaks of $Pr^{3+}$ from (a), respectively.

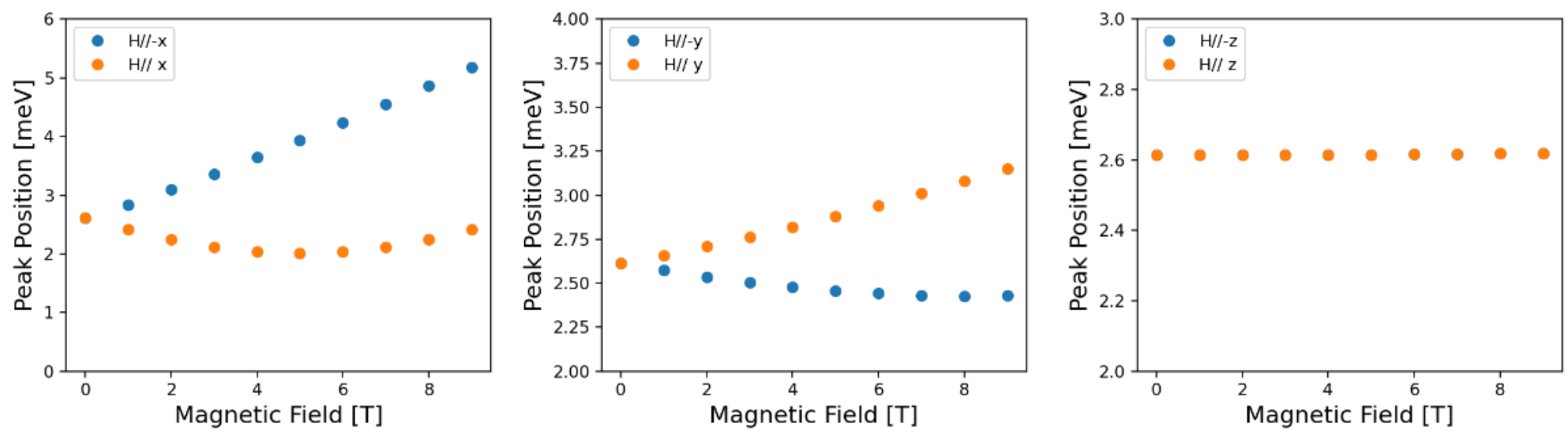

**Fig. 24** (a), (b), and (c) show the field dependence of the split ground-state CF excitation energy of $Pr^{3+}$ in $PrFeO_3$ under external fields applied in the same and opposite directions as the 10 T internal field along the *x*-, *y*-, and *z*-directions, respectively.

### 3.2.2 $HoFeO_3$

$Ho^{3+}$ is another non-Kramers rare earth ions which has 10 electrons in the 4*f* orbital. The total spin number is 2 and the total angular momentum *J* is 8. Thus, the ground state of $Ho^{3+}$ in $HoFeO_3$ can be denoted as $^5I_8$. In a compound, this ground state of $Ho^{3+}$ suffers from the CF effect and splits into $2J+1=17$ individual singlet sublevels. The energy degeneracies of CF levels are highly dependent on the local symmetry. The CF excitations of $Ho^{3+}$ in $HoFeO_3$ have been measured by A. K. Ovsyanikov et al.[24] and G. A. Stewart et al.[25] using inelastic neutron scattering technique. The CF excitation levels of $Ho^{3+}$ in $HoFeO_3$ were simulated by O. V. Usmanov et al.[33] using the point-charge model. However, the simulated CF levels did not agree well with the excitation peaks reported by A. K. Ovsyanikov et al.[24] From the reported inelastic neutron measurements,[24] the following CF excitation peaks were confirmed for $HoFeO_3$: ~ 10 meV, ~ 15 meV, ~ 20 meV, ~ 29.5 meV, and ~ 45 meV, see Table 10. Additionally, a low-energy excitation peak between 0 and 1 meV was observed at temperatures below 60K, showing a clear temperature dependence in this range. These findings suggest that CF excitations in $HoFeO_3$ are highly complex. Understanding these phenomena remains an open question, warranting further investigation for deeper insight.

Using the point-charge model, we calculated the CF excitations of $Ho^{3+}$ in $HoFeO_3$ and fitted the model to the reported CF excitation peaks in Table 10, yielding the parameters listed in Table 11. With these fitted parameters, we simulated CF excitations under various internal magnetic fields. The results for internal fields are shown in Figs. 25, 26, and 27. A key distinction in $HoFeO_3$ is the intense ground-state CF excitation at zero energy transfer, in contrast to the non-Kramers system $PrFeO_3$, where no such excitation appears under zero internal or external fields. This difference likely arises from the greater number of singlet levels in $HoFeO_3$ compared to $PrFeO_3$, increasing the likelihood of forming pseudo-doublet CF levels due to accidental degeneracies. Table 12 lists low-energy CF eigenvalues for $HoFeO_3$, revealing nearly degenerate singlet pair states at ~ 0 meV, ~ 10 meV, and ~ 29.6 meV. These closely spaced pairs can be considered pseudo-doublets, suggesting that the strong ground-state excitation in $HoFeO_3$ originates from a pseudo-doublet ground state.

**Table 10**. CF excitation levels of $HoFeO_3$ [34]

| Peak No. | 1 | 2 | 3 | 4 |
|---|---|---|---|---|
| Energy (meV) | 10 | 15 | 20 | 29.5 |

**Table 11.** CF parameters of $Ho^{3+}$ by fitting to the peaks and intensities in Table 10

| Param | $B_2^0$ | $B_2^2$ | $B_2^{-2}$ | $B_4^0$ | $B_4^2$ | $B_4^{-2}$ |
|---|---|---|---|---|---|---|
| Value | 0.14192 | 0.15816 | -0.19008 | -3.8629e-05 | -3.1776e-03 | -2.8690e-03 |
| Param | $B_4^4$ | $B_4^{-4}$ | $B_6^0$ | $B_6^2$ | $B_6^{-2}$ | $B_6^4$ |
| Value | -1.5841e-03 | -1.3646e-03 | -8.8670e-06 | -2.8895e-05 | 2.1902e-05 | -2.5154e-05 |
| Param | $B_6^{-4}$ | $B_6^6$ | $B_6^{-6}$ | | | |
| Value | 2.0911e-05 | -5.2100e-05 | -4.1649e-05 | | | |

**Table 12.** Low-energy eigenvalues of $Ho^{3+}$ CF levels in $HoFeO_3$ calculated from the model in the text

| Peak No. | 1 | 2 | 3 | 4 | 5 | 6 | 7 | 8 | 9 |
|---|---|---|---|---|---|---|---|---|---|
| Energy [meV] | 0.0 | 0.03193 | 10.4091 | 10.70134 | 15.5752 | 20.5395 | 26.70545 | 28.7299 | 29.5545 |

Internal magnetic fields along the *x*-direction slightly split both the ground state and the first excited state at 10 meV, with new peaks appearing as shoulders on the high-energy side. These fields also shift the excitation peaks at 15 meV and 20 meV to higher energies without splitting them. As previously noted, the excitations at 0 meV and 10 meV correspond to pseudo-doublet states, which are split by the internal fields. In contrast, the peaks at 15 meV and 20 meV are singlets, which magnetic fields can only shift rather than split.

Internal fields along the *y*-direction produce similar but significantly stronger effects on these CF peaks. Under *y*-direction fields, the split peaks at approximately 0 meV and 10 meV are well separated from the main peaks. Additionally, the energy shifts of the excitations at 15 meV and 20 meV are much larger than those observed under *x*-direction fields.

The internal magnetic fields along the *z*-axis produce effects entirely distinct from those described above. No splitting is observed for the ground state peak or the excitation peak at 10 meV, and no energy shifts occur for the ground state peak or the excited states at 15 meV and 20 meV. The peak at 10 meV exhibits only a slight shift. However, the peak at approximately 29.5 meV splits under the field along this direction, consistent with the theoretically predicted pseudo-doublet at this energy. In contrast, this same peak shows no splitting under fields along the *x*- or *y*-directions.

As shown in Figs. S10 ~ S12 in the Supplemental Material, when the external magnetic fields are applied, the impacts of the fields along the different directions are like what were observed in the internal magnetic fields. However, the effects are much stronger along all three different directions at the same level of the fields. Especially, along the *y* direction, the excited states at 15 meV and 20 meV are significantly shifted by the external fields. The split peak from 10 meV pseudo-doublet have crossed the original 15meV peak while the 15meV peak shifts to ~20.5 meV and the 20 meV peak shifts to ~26.5 meV at 9T. These peak shifts were the largest observed in all the $REFeO_3$ discussed in this work.

The combined effects of internal and external magnetic fields were simulated by fixing the internal magnetic field at 10 T and varying the external magnetic fields along the *x*-, *y*-, and *z*-directions. The results for the *y*-direction are presented in Figs. 28 and 29. We observed that external magnetic fields applied in the opposite direction to the internal field partially cancel its effects, whereas those aligned with the internal field amplify the shifting and splitting of the CF excitation peaks. The field dependence of the external magnetic fields along the *x*-, *y*-, and *z*-directions is depicted in Fig. 30. Notably, the split excitation peak from the ground state exhibits field dependence along the *x*- and *y*-directions but none along the *z*-direction, with the dependence along the y-direction being significantly stronger than along the *x*-direction. The 10 T internal magnetic field is somehow compensated by an external field of 1.7 T along the *y*-direction, whereas along the *x*-direction, the same effect requires an external field of 4 T. Additionally, the slope of the field dependence along the *y*-direction is markedly steeper than that along the x-direction.

It is interesting to point it out that all the Kramers ions ($Nd^{3+}$, $Er^{3+}$, and $Yb^{3+}$) in $REFeO_3$ have the same compensating external magnetic fields along the three directions, while all the non-Kramers ions ($Pr^{3+}$ and $Ho^{3+}$) in $REFeO_3$ have very different compensated external magnetic fields along the *x*, *y*, and *z* directions according to the calculated results presented above. This could be an intrinsic difference between the degenerated doublets in the Kramers systems and the non-degenerated (or accidentally-degenerated) singlets in the non-Kramers systems.

Inelastic neutron scattering measurements of $HoFeO_3$ reveal a low-energy excitation below 1 meV,[25] which we attribute to one of the pseudo-doublet ground state peaks. This peak shifts to higher energy due to internal magnetic fields from the $Fe^{3+}$ and $Ho^{3+}$ sublattices. A similar energy shift, documented in Ref[25], exhibits two distinct steps: one between approximately 50 K and 60 K, and another below 10 K. The high-temperature shift, from 0.3 meV to 0.5 meV, corresponds to the spin reorientation transition of the $Fe^{3+}$ magnetic sublattice, known to occur between 50 K and 60 K. In contrast, the low-temperature shift below 10 K likely results from an increasing internal magnetic field from the $Ho^{3+}$ sublattice, which approaches long-range magnetic order as the temperature decreases from 10 K to 1 K. Thus, our CF model effectively accounts for these observations.

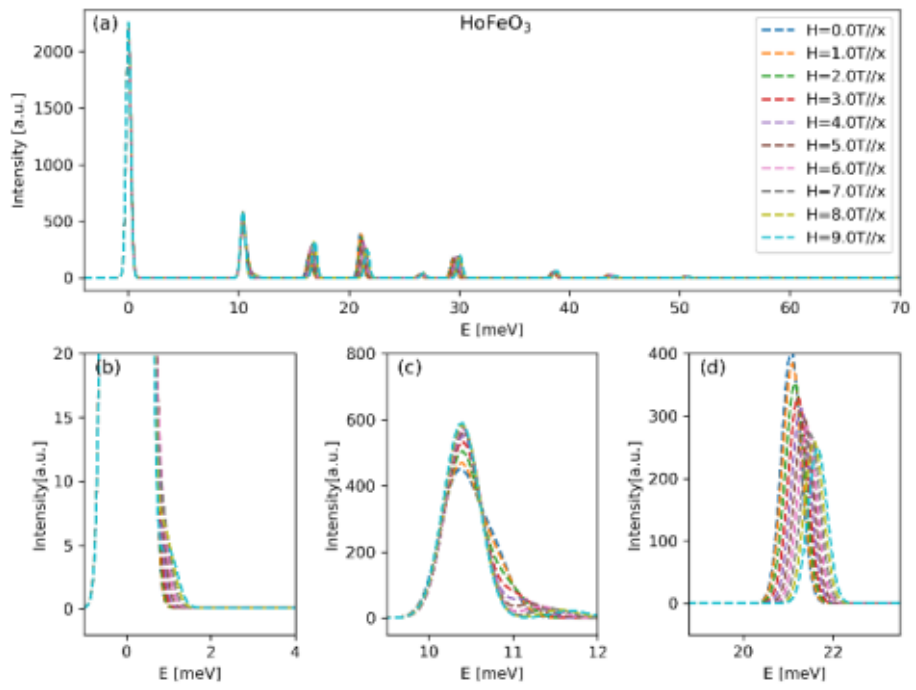

**Fig. 25** (a) CF excitation spectra of $Ho^{3+}$ in $HoFeO_3$ at different internal magnetic fields along the *x*-direction. (b), (c), and (d) show zoomed-in views of the first, second, and third CF excitation peaks of $Ho^{3+}$ from (a), respectively.

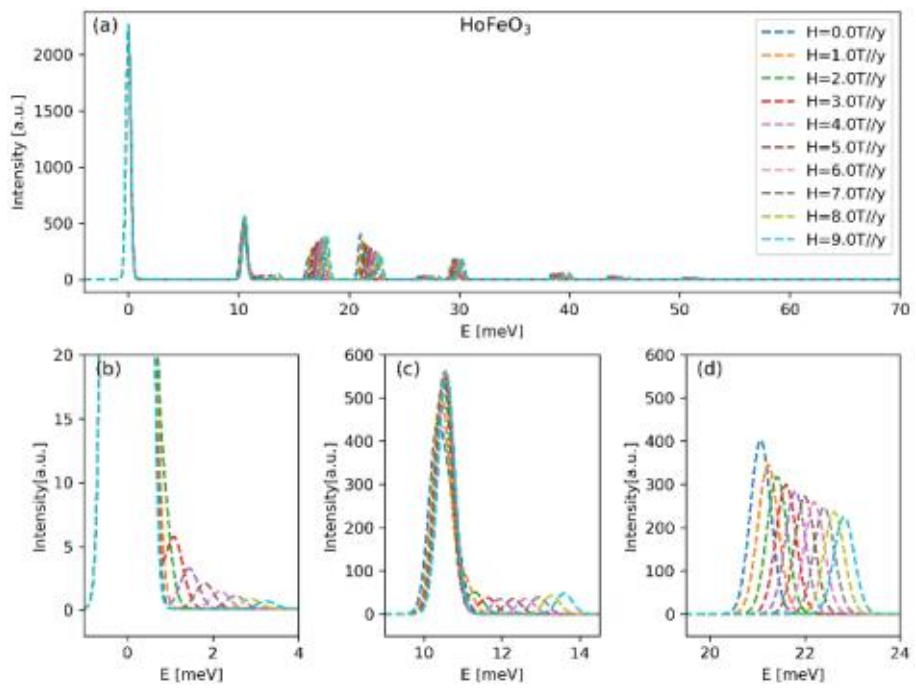

**Fig. 26** (a) CF excitation spectra of $Ho^{3+}$ in $HoFeO_3$ at different internal magnetic fields along the *y*-direction. (b), (c), and (d) show zoomed-in views of the first, second, and third CF excitation peaks of $Ho^{3+}$ from (a), respectively.

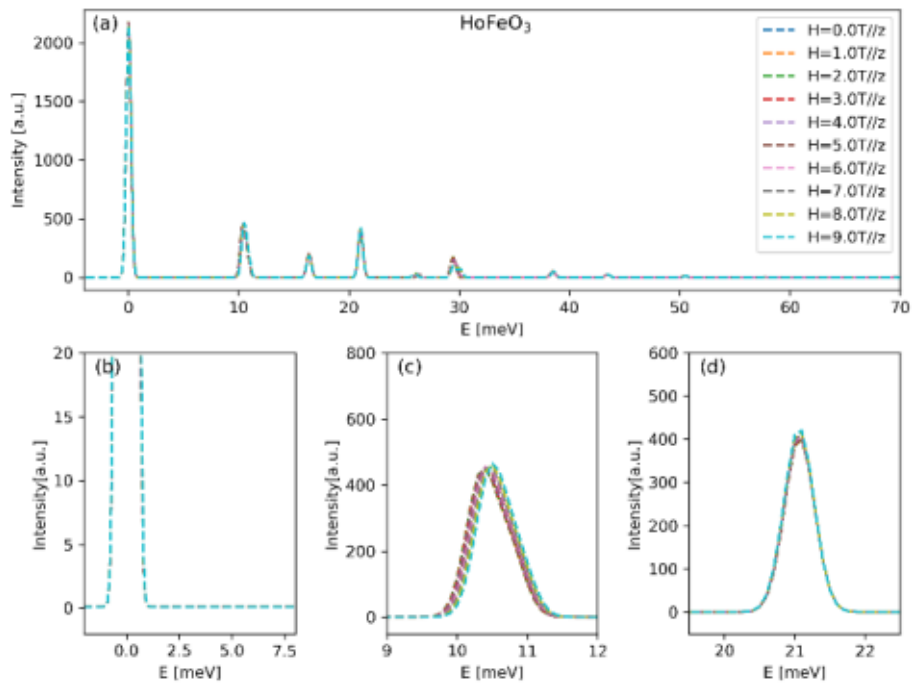

**Fig. 27** (a) CF excitation spectra of $Ho^{3+}$ in $HoFeO_3$ at different internal magnetic fields along the *z*-direction. (b), (c), and (d) show zoomed-in views of the first, second, and third CF excitation peaks of $Ho^{3+}$ from (a), respectively.

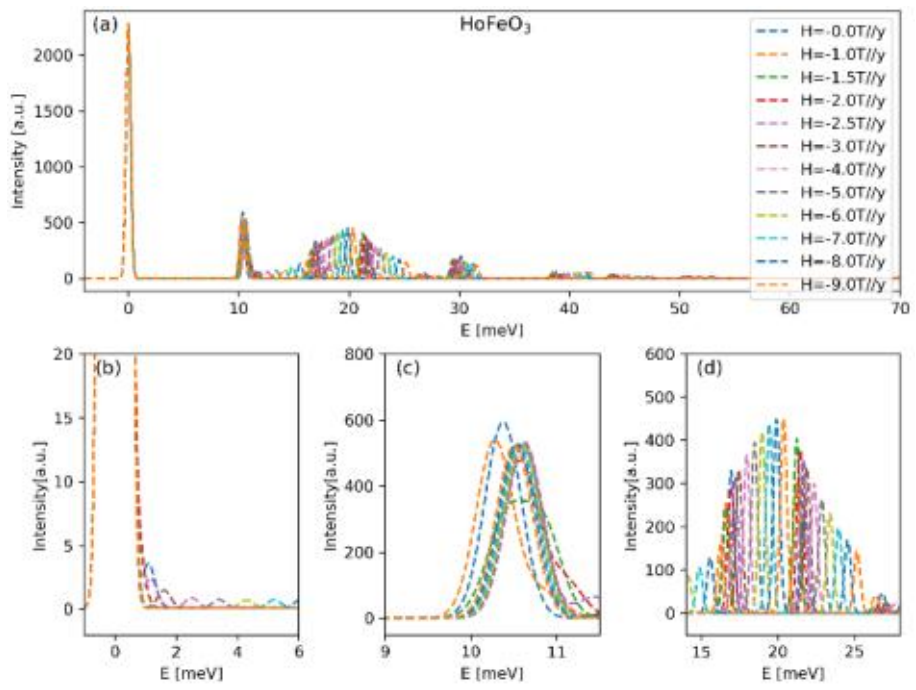

**Fig. 28** (a) CF excitation spectra of $Ho^{3+}$ in $HoFeO_3$ at different external fields along the *opposite* direction, assuming a 10 T internal field along the *y* direction. (b), (c), and (d) show zoomed-in views of first, second, and third CF peaks of $Ho^{3+}$ from (a), respectively.

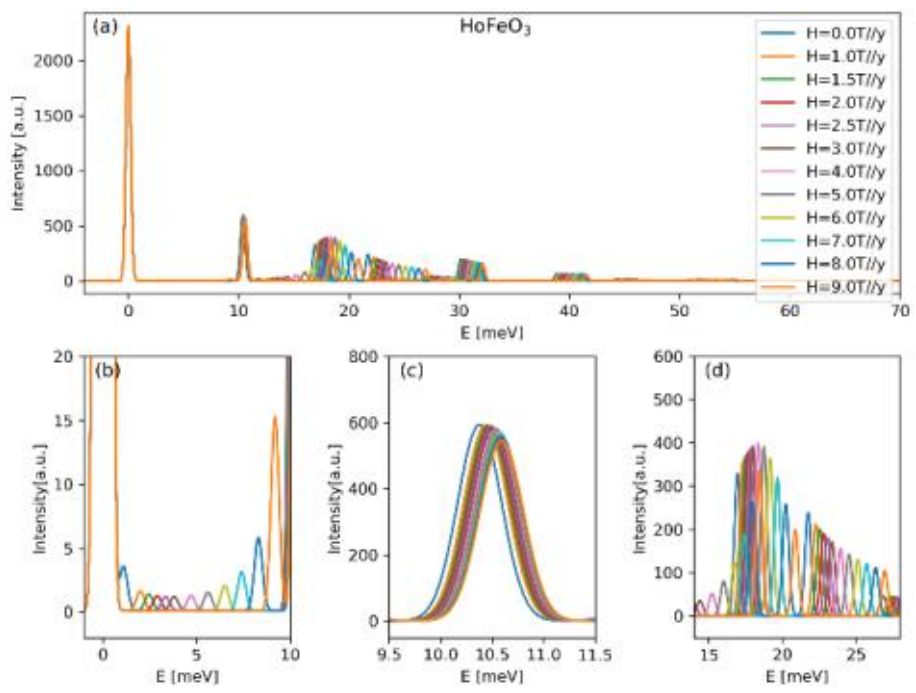

**Fig. 29** (a) CF excitation spectra of $Ho^{3+}$ in $HoFeO_3$ at different external fields along the *same* direction, assuming a 10 T internal field along the *y* direction. (b), (c), and (d) show zoomed-in views of first, second, and third CF peaks of $Ho^{3+}$ from (a), respectively.

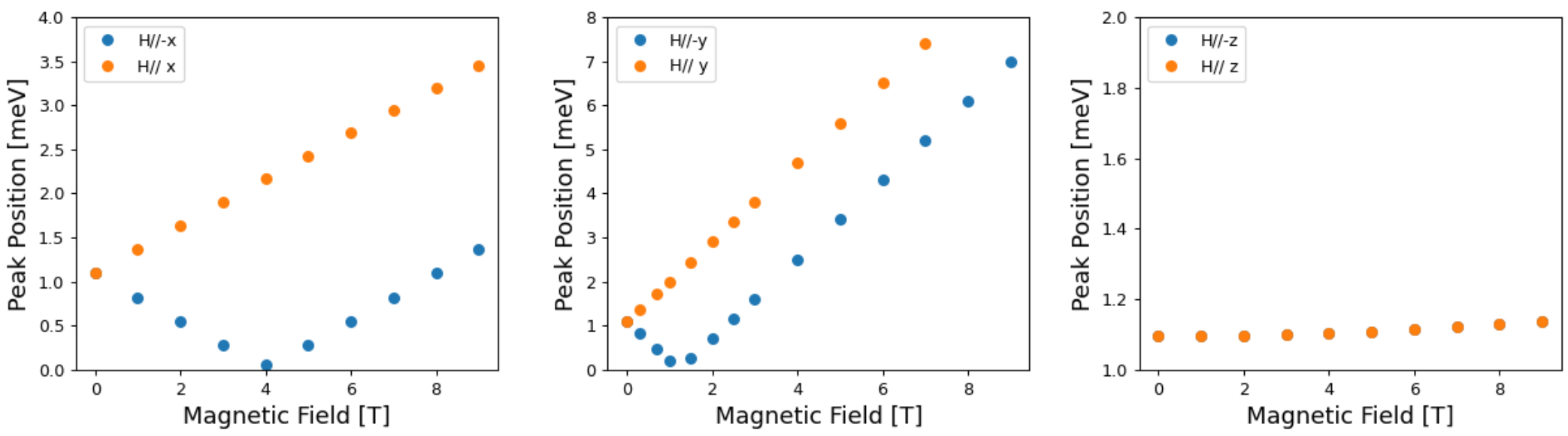

**Fig. 30** (a), (b), and (c) show the field dependence of the split ground-state CF excitation energy of $Ho^{3+}$ in $HoFeO_3$ under external fields applied in the same and opposite directions as the 10 T internal field along the *x*-, *y*-, and *z*-directions, respectively.

3.3 *Discussion*

The CF excitations in the low-energy range were reported in $NdFeO_3$,[23] $ErFeO_3$,[11] $YbFeO_3$,[8] $NdGaO_3$,[35] and $HoFeO_3$[24,25] et al. It seems a quite universal phenomena in this series of RE-TMOs. However, similar effects have not been reported or widely observed in other systems. Some tentative interpretations to these low-energy excitations were reported previously. One explanation to the low-energy excitation at ~0.17 meV in $NdGaO_3$ previously attribute the excitation to the transition between hyperfine-split nuclear level of the $^{143}Nd$ and $^{145}Nd$ isotopes with spin I = 7/2.[35] However, the paper also pointed that only one excitation peak was observed even though two excitation peaks were expected because of the two isotopes. This explanation basically ignored the fact the excitation peak is highly correlated with the ordering of the $Nd^{3+}$ magnetic moments. $Nd^{3+}$ magnetic moments form a long-range antiferromagnetic ordering at ~ 1 K. The excitation peak showed up below this temperature as well. This strongly supports that the excitation is strongly correlated to electronic excitations rather than nuclear excitations. A similar low-energy excitation, peaking at 0.49 meV, was observed in $NdFeO_3$. The investigators proposed that this excitation arises from the splitting of the $Nd^{3+}$ ground state doublet in the CF spectrum, driven by the internal molecular field. However, the original paper provided no direct calculations or theoretical simulations to substantiate this hypothesis. Another case study involves $ErFeO_3$, where inelastic neutron scattering revealed a low-energy, non-dispersive excitation ranging from approximately 0.37 meV to 0.75 meV. CF modelling demonstrates that this low-energy excitation can be effectively explained by the internal magnetic field scenario. Despite the difficulty in quantitatively assessing the energy accuracy of the Stevens-operator model fits for $NdFeO_3$ and $ErFeO_3$, the strong overall agreement between experimental observations and modelled behaviour lends strong confidence to this scenario. Nevertheless, several critical questions remain unresolved. Is this low-energy excitation a universal phenomenon across all rare earth magnets? Given that $Er^{3+}$ and $Nd^{3+}$ are Kramers ions, could similar effects be observed in other rare earth orthoferrites with non-Kramers ions? Furthermore, can analogous effects occur in other rare earth transition metal compounds such as oxides (RE-TMOs), chlorides, and fluorides, irrespective of the local symmetry of the rare earth ions? These questions are both intriguing and significant for further investigation.

Previously, B. Z. Malkin had studied the electron-phonon coupling impact to CF excitations in crystalline materials.[16] More recently, B. Z. Malkin et. al.[22] carefully studied how lattice deformations induced by impurities and defects split the CF excitation peaks in the $Yb^{3+}$, $Sm^{3+}$, $Er^{3+}$ doped $Cs_2NaYF_6$ and $Cs_2NaScF_6$ after X. Zhou's work in $Cs_2NaLnCl_6$ (Ln = $Er^{3+}$ and $Yb^{3+}$)[37]. First, we must admit that electron-phonon coupling exists in $REFeO_3$ undoubtedly and play roles in the CF excitation in some special conditions. However, it is well-known that CF-phonon coupling takes place in the region where the phonon and CF energies are close to each other. Phonon excitations are significantly suppressed at low temperature. Thus, the phonon impact on the measured result is much limited at the measuring temperatures of those $REFeO_3$ samples mentioned in the current study. Phonon is stronger at larger Q while CF excitation is stronger at lower Q due to its magnetic nature. Thus, most of the CF spectra were measured in the low Q region, where phonon structure factors are very low. From the experimental point of view at low temperature and low Q range, therefore, the reported or measured CF spectra have very limited impacts from phonon.

Second, we should consider the impurity and defects in the $REFeO_3$ crystals, which could cause the CF splitting. In the case of Kramers ions such as $Yb^{3+}$, $Er^{3+}$, and $Nd^{3+}$, their ground states of their CF are not sensitive to any local structural deviation such as local distortions induced by the strains and defects because Kramers ions intrinsically have a degenerate doublet ground state no matter the local symmetry of the site. For the non-Kramers ions, the local distortion effect could be more complicated. However, several experimental evidence clearly exclude the structural origin of the energy shifts observed in $REFeO_3$. As shown in $ErFeO_3$, the low-energy CF excitation shifts significantly upon cooling through the Néel temperature $T_N$ of $Er^{3+}$ ions, which indicates it strongly couple to the magnetic order of $Er^{3+}$ spin. There is no structural phase transition at this temperature range. The reported low-energy CF excitation in $YbCl_3$[36] shows the similar features as observed in $YbFeO_3$. Considering the different local symmetries of $Yb^{3+}$ in$YbFeO_3$[30] and $YbCl_3$[36], this means that the ground state splitting of $Yb^{3+}$ is not symmetry dependent. One possible explanation to this widely observed excitation is that the internal magnetic field induced by the exchange interaction within the $Yb^{3+}$ sublattice at low temperature. Such internal magnetic fields can lift the degeneracy of the doublet of the CF excitation levels in Kramers ions like $Yb^{3+}$ and $Er^{3+}$.

For non-Kramers ions, the situation becomes more complicated. First, the existence of pseudo-doublets is highly dependent on the RE ion itself and the local symmetry. Structural distortions could induce a pseudo-doublet splitting. Using the same reasons, low temperature and low Q measurements, we could also exclude the possibility that phonon induces the ground state splitting in these non-Kramers. The ground state splitting of CF excitations of non-Kramers $REFeO_3$ is hard to predict because it is hard to know if the system has a pseudo-doublet as its ground state.

This study examines the behaviour of both Kramers and non-Kramers rare earth ions in $REFeO_3$. Our simulations reveal that ground state splitting is consistently observed in Kramers rare earth ions, owing to their intrinsic doublet ground states. In contrast, such splitting is typically absent in non-Kramers rare earth ions, which possess intrinsic singlet ground states. However, if accidental degeneracy occurs among the singlet ground states of non-Kramers ions, a pseudo-doublet state may form. In such cases, splitting of this pseudo-doublet state can be anticipated in non-Kramers $REFeO_3$ systems. The local symmetry plays important role in the CF excitation of non-Kramers $REFeO_3$ systems because the high local symmetry could induce degeneracy of CF levels.

Collectively, this study demonstrates that a degenerate ground state—whether a true doublet or a pseudo-doublet—combined with low local symmetry and internal magnetic fields, is crucial for the emergence of low-energy excitation states in RE-TMOs. These excitations manifest as split peaks from the ground state doublet, induced by internal fields. Additionally, this study also emphasizes the need to be cautious when interpreting all low-energy excitations in rare-earth candidate compounds as the so-called fingerprint of QSL excitations. In some cases, the ground state splitting, especially, in Kramers rare earth compounds, could induce a low-energy excitation spectra around 1meV even without long range magnetic ordering. Furthermore, such low-energy excitation could demonstrate a very weak dispersion due to the magnetic exciton effect. Experimentally excluding such possibilities become essential to correctly draw conclusion if one magnetic system is a QSL system.

After all the analysis and simulation work had been finished, we drew our attention to the CF software SPECTRE.[38] It is different from the CF functions in Mantid and the Python package PyCrystalField since it builds on the Wybourne tensor operator formalism rather than the Stevens operator-equivalent formalism. The Wybourne operators are based on the spherical tensor operators while the Stevens operators are based on real tesseral harmonics and angular momentum components. This means that the CF parameters in the Wybourne operator method may reflect the crystal environmental contribution directly without considering the contribution of the angular momentum components. In the current study, using the Wybourne operator method may provide more information about the common feature of the $C_s$ symmetry of the 4*c* sites in $REFeO_3$ no matter which RE ions occupy this site.

A continuation of the current study with the Wybourne operator method is worthwhile to be carried out in future from this point of view.

## 3. Conclusion

Using previously reported CF excitation data, we developed CF excitation models for Kramers ions ($Nd^{3+}$, $Er^{3+}$, $Yb^{3+}$) and non-Kramers ions ($Pr^{3+}$, $Ho^{3+}$) in $REFeO_3$ orthoferrites. These models enabled a systematic investigation of the effects of internal and external magnetic fields, yielding numerous insightful findings. Our analysis reveals that the ground CF states of Kramers ions in $REFeO_3$ consistently split under internal or external magnetic fields, a consequence of their intrinsically degenerate doublet ground states. In contrast, the ground states of non-Kramers ions, such as $Pr^{3+}$ in $REFeO_3$, remain unsplit under similar conditions due to their non-degenerate singlet nature. However, in the case of $Ho^{3+}$, the ground state forms an accidentally degenerate pseudo-doublet—comprising two singlet states with nearly identical energies—that also splits under internal or external fields. Furthermore, our simulations highlight a pronounced anisotropy in the response of CF energy levels to magnetic fields along the *x*-, *y*-, and *z*-directions. Conclusively, these results suggest that the widely observed low-energy excitations (< 1 meV) in $REFeO_3$ and $REGaO_3$ could arise from the splitting of CF ground states, induced by internal magnetic fields from the $RE^{3+}$ sublattice, the $Fe^{3+}$ sublattice, or both.

## Acknowledgement

We acknowledge the support of the Australian Centre for Neutron Scattering, Australian Nuclear Science and Technology Organisation, and the Australian Government through the National Collaborative Research Infrastructure Strategy, in supporting the neutron research infrastructure. We would like to thank ACNS for allocating neutron beamtime for the experiment on the thermal-neutron triple-axis spectrometer Taipan via the proposal P3847.

## Declaration of generative AI and AI-assisted technologies in the writing process

During the preparation of this work the author used ChatGPT to check grammar and improve the clarity and accuracy of some parts of the text. After using this tool, the author reviewed and edited the content as needed and take full responsibility for the content of the publication.

# Supplementary Material for "Internal and External Field Effects upon Crystal Field Excitations in $REFeO_3$ (RE = $Nd^{3+}$, $Er^{3+}$, $Yb^{3+}$, $Pr^{3+}$, and $Ho^{3+}$)"

Guochu Deng*

Australian Centre for Neutron Scattering, Australian Nuclear Science and Technology Organisation, New Illawarra Road, Lucas Heights NSW 2234, Australia

## 1. Irreducible Representation Analysis

According to the group theory analysis, the double group $C_s$ has only two symmetry elements: one identity element E and one reflection plane element ($\sigma_h$). Thus, the character table of $C_s$ is listed in Table S1. There are four irreducible representations in total: Γ1 ⊕ Γ2 ⊕ Γ3 ⊕ Γ4. The first two Γ1 and Γ2 are one-dimensional (1D), corresponding to the non-Kramers rare earth ions. The last two Γ3 and Γ4 are two-dimensional (2D), corresponding to the Kramers rare earth ions.

**Table S1.** The character table of the double group $C_s$

| $C_s$ | $E$ | $\sigma_h$ | $E$ | $\sigma_h$ | Type |
|---|---|---|---|---|---|
| Γ1 | 1 | 1 | 1 | 1 | Non-Kramers |
| Γ2 | 1 | -1 | 1 | -1 | Non-Kramers |
| Γ3 | 2 | 0 | -2 | 0 | Kramers |
| Γ4 | 2 | 0 | -2 | 0 | Kramers |

## 2. Simulation Results

### 2.1 *Kramers Ions in $REFeO_3$*

2.1.1 $NdFeO_3$

Besides the simulated internal effects on the crystal field (CF) excitations of $NdFeO_3$ in the main article, the external field effects along the *x*, *y*, and *z* directions were studied through simulations as well. The results were presented in Figs. S1, S2, and S3, respectively. When applying external magnetic fields, it seems that the fields have similar impacts to the individual excitation peaks as what the internal magnetic fields do. However, the energy shifts are more significant with the external fields in the current case. It is interesting to compare the energy shifts along the three directions. The magnetic fields along the *y* direction have the strongest impacts in shifting the ground state and third excitation peaks. When the internal field changes into the *z* direction, the impacts to these CF peaks are significantly different from the effects observed for the *x* and *y* directions. First, the split new peaks of the ground state are much more intense in the fields along *z*. Second, the first excitation peak obviously shifts to the high energy side, significantly different from the tiny shifts observed for the *x* and *y* directions. Furthermore, the third excitation peak near 45 meV shows little splitting, in contrast to the apparent splitting in the fields along the *x* and *y* directions. Considering the external field impacts on the CF excitations in $NdFeO_3$, the impacts of the magnetic fields along the three directions are very similar to the cases of the internal fields. Comparing them one by one, we could find that the external fields have slightly stronger impacts from the energy shift point of view.

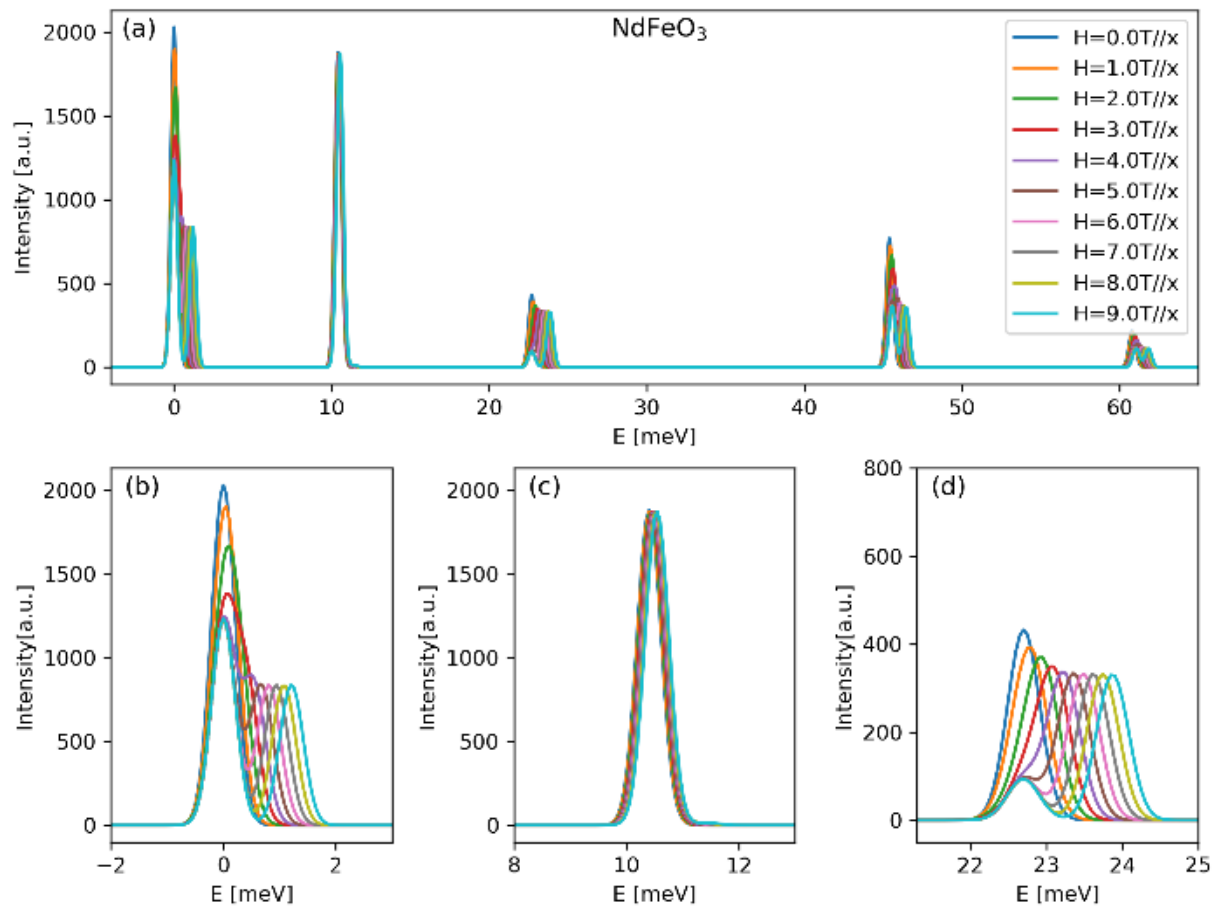

**Fig. S1** (a) CF excitation spectra of $Nd^{3+}$ in $NdFeO_3$ at different external magnetic fields along the *x*-direction. (b), (c), and (d) show zoomed-in views of the first, second, and third CF excitation peaks of $Nd^{3+}$ from (a), respectively.

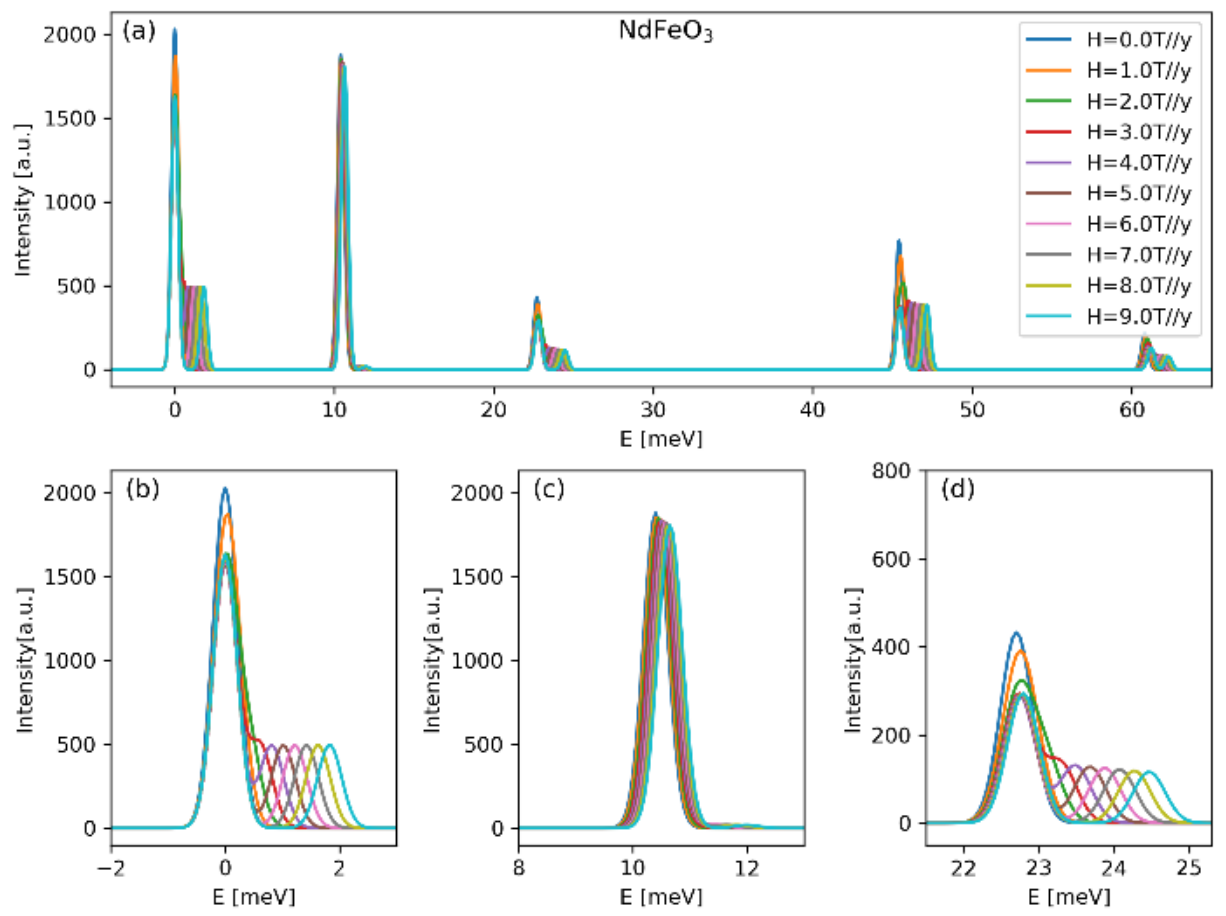

**Fig. S2** (a) CF excitation spectra of $Nd^{3+}$ in $NdFeO_3$ at different external magnetic fields along the *y*-direction. (b), (c), and (d) show zoomed-in views of the first, second, and third CF excitation peaks of $Nd^{3+}$ from (a), respectively.

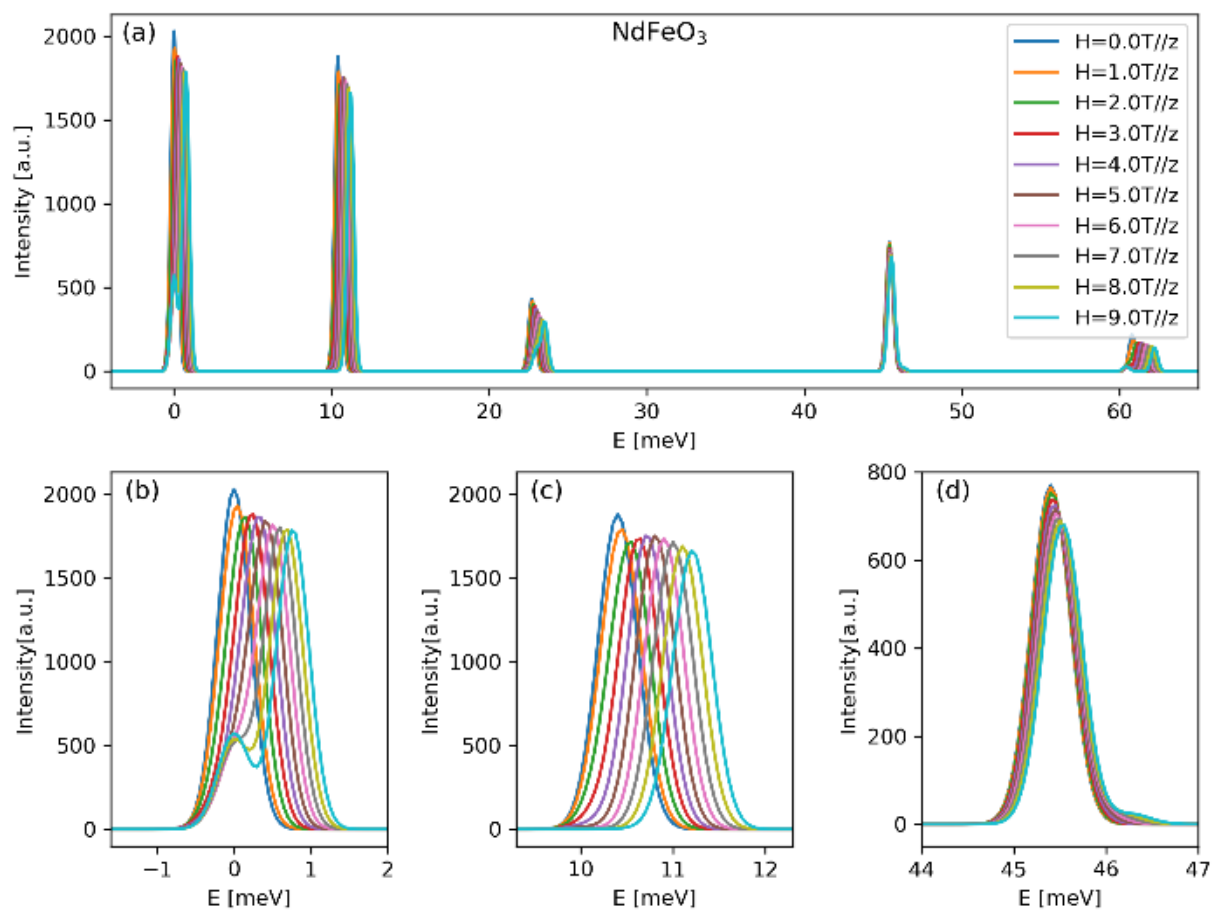

**Fig. S3** (a) CF excitation spectra of $Nd^{3+}$ in $NdFeO_3$ at different external magnetic fields along the $z$-direction. (b), (c), and (d) show zoomed-in views of the first, second, and third CF excitation peaks of $Nd^{3+}$ from (a), respectively.

### 2.1.2 $YbFeO_3$

We also studied the impacts of external magnetic fields upon the CF excitations of $YbFeO_3$. Figs. S4, S5, and S6 show the simulated CF excitation spectra of $Yb^{3+}$ in $YbFeO_3$ under the external magnetic fields. In general, the external magnetic fields along all the directions have much stronger impacts than the internal fields at the same level. The external magnetic fields along the $x$ and $y$ directions clearly split all the CF peaks. All the new peaks significantly shift to the higher energy side. However, the external magnetic fields along the $z$ direction only split the ground state peak and strongly shifts the second and the third peaks to higher energy without splitting. These observations are similar to impacts induced by the internal magnetic fields.

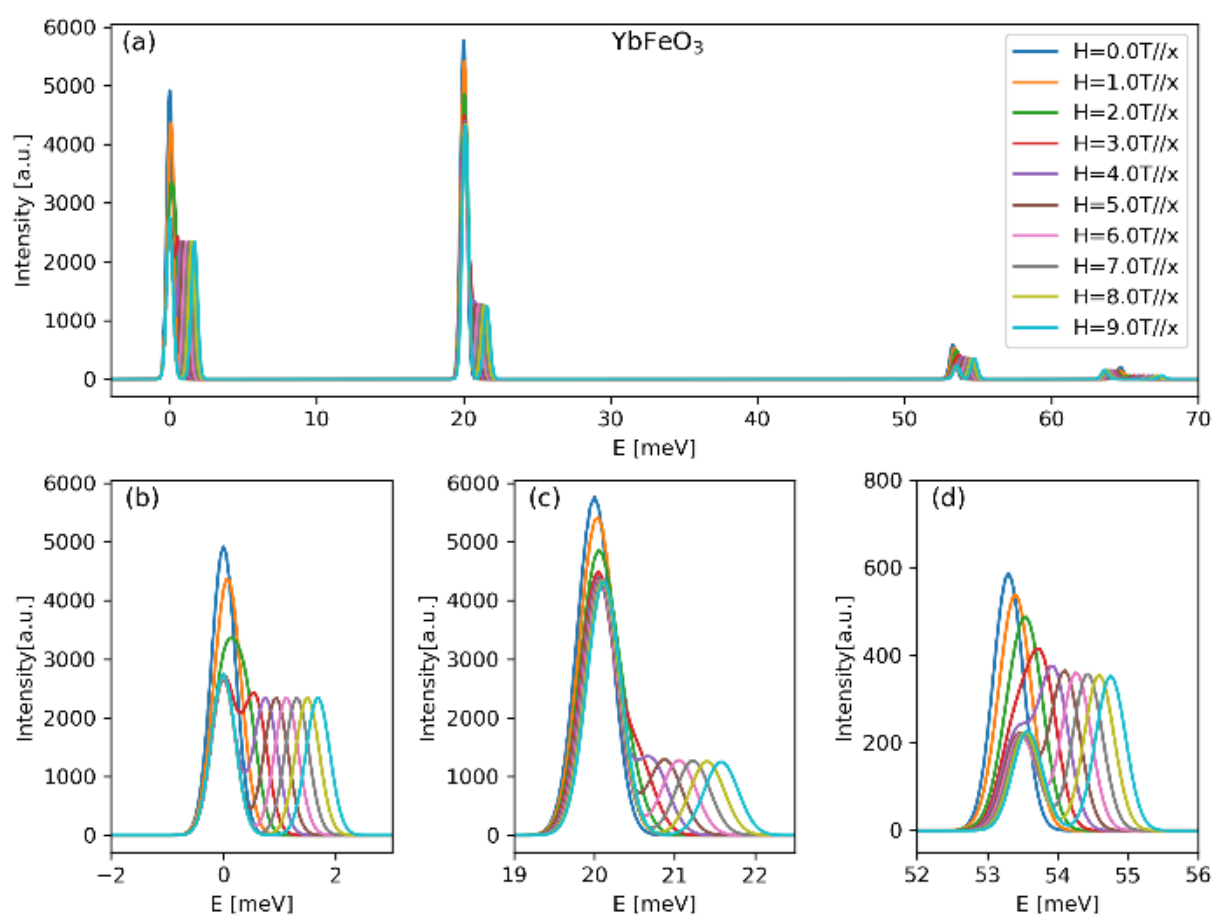

**Fig. S4** (a) CF excitation spectra of $Yb^{3+}$ in $YbFeO_3$ at different external magnetic fields along the $x$-direction. (b), (c), and (d) show zoomed-in views of the first, second, and third CF excitation peaks of $Yb^{3+}$ from (a), respectively.

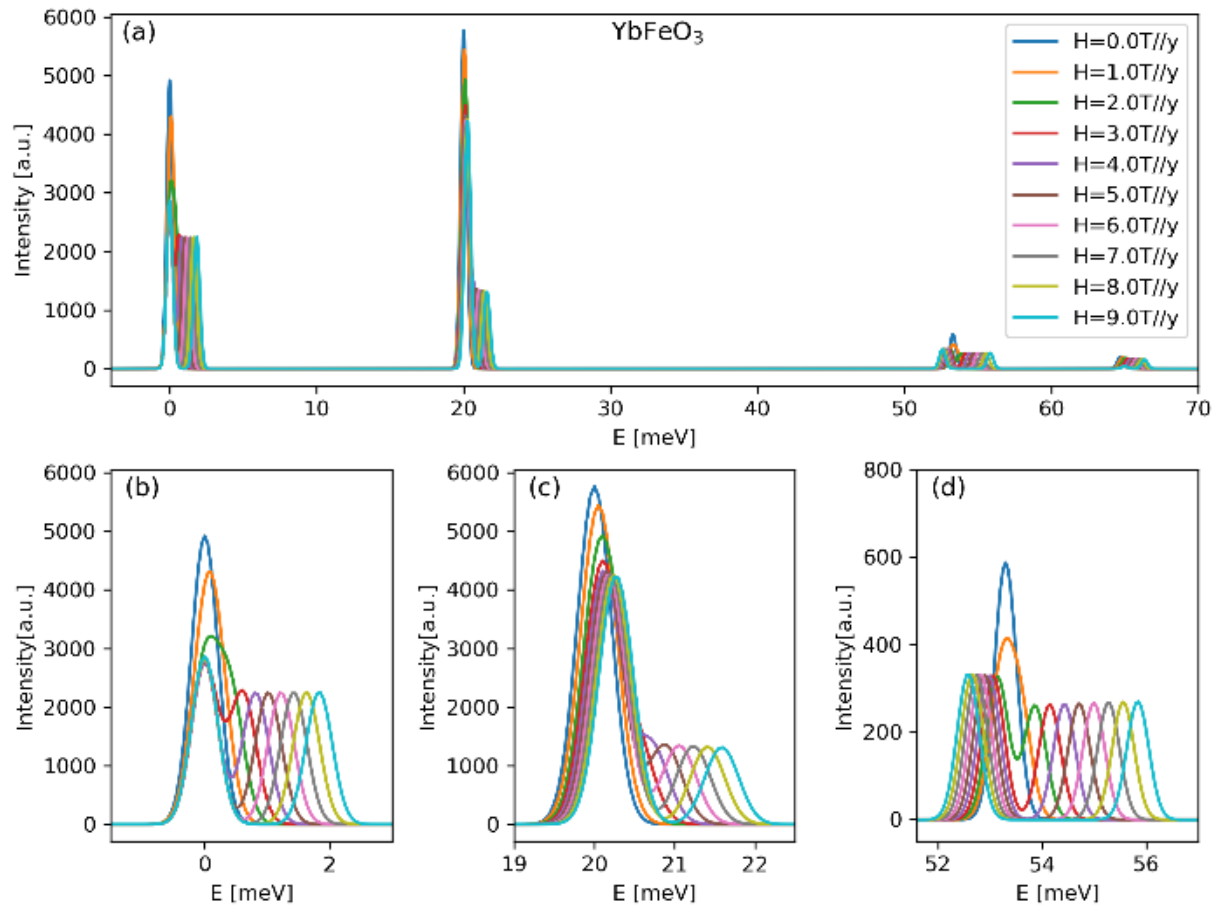

**Fig. S5** (a) CF excitation spectra of $Yb^{3+}$ in $YbFeO_3$ at different external magnetic fields along the *y*-direction. (b), (c), and (d) show zoomed-in views of the first, second, and third CF excitation peaks of $Yb^{3+}$ from (a), respectively.

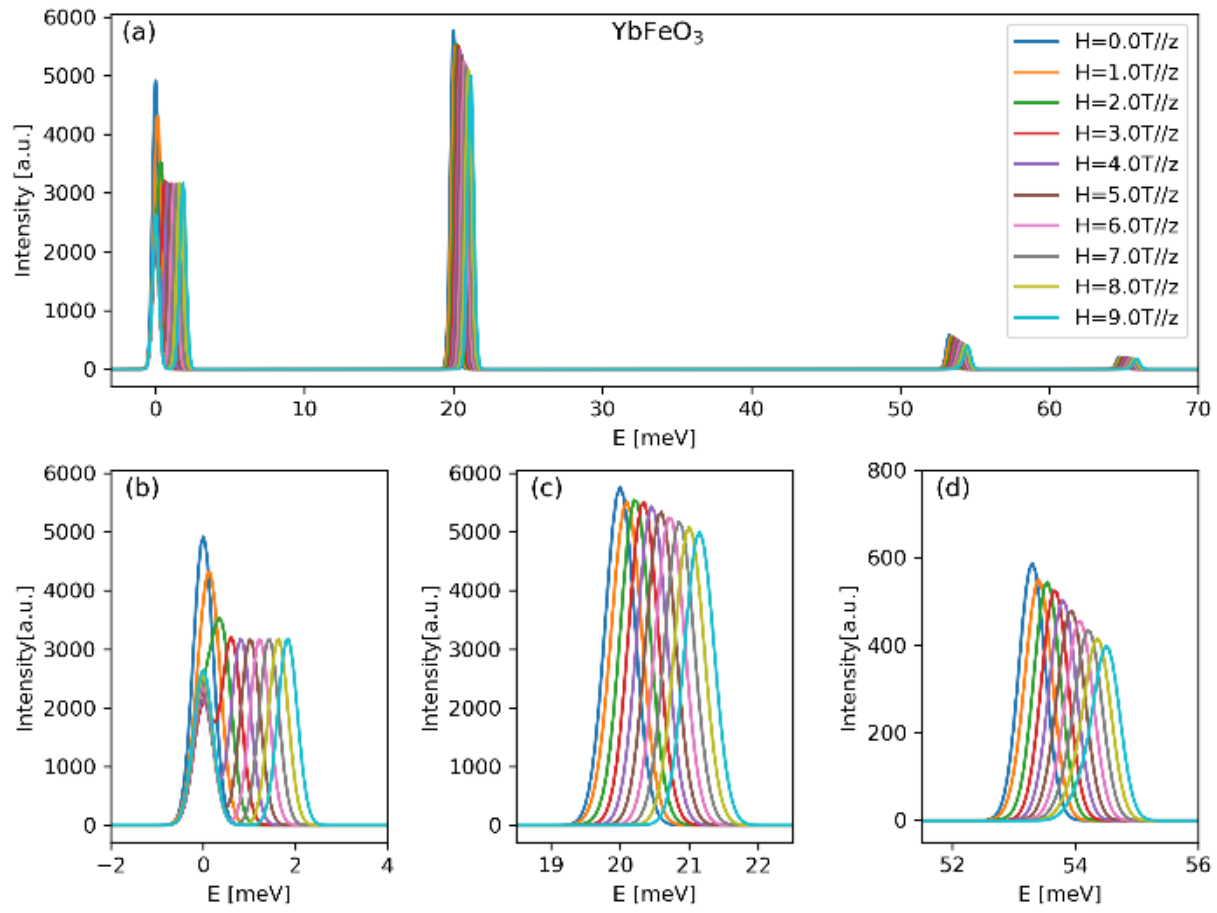

**Fig. S6** (a) CF excitation spectra of $Yb^{3+}$ in $YbFeO_3$ at different external magnetic fields along the *z*-direction. (b), (c), and (d) show zoomed-in views of the first, second, and third CF excitation peaks of $Yb^{3+}$ from (a), respectively.

**2.2** ***Non-Kramers ions in $REFeO_3$***

2.2.1 $PrFeO_3$

The external magnetic field effects were simulated for $PrFeO_3$ using the same model shown in the main article. The simulated results are shown in Figs. S7, S8, and S9. In general, the external magnetic fields demonstrate similar, but stronger, effects as the internal fields. These figures clearly demonstrate that the external magnetic fields have a much stronger impact comparing to the internal fields at the same levels. In Fig. S7, the external fields along the *x* direction induce the ground state excitation, which is about three times stronger than the intensities induced by the internal fields in Fig. 19 in the main article. However, the peak at 0.87 meV does not show many changes in the peak position and intensity when comparing to the previous results shown in Fig. 19. The rest excitation peaks at higher energies demonstrate higher intensities and larger energy shifts in the external fields than in the internal fields. The similar enhancements are observed for the external fields along the *y* direction, as shown in Fig. S8. The external fields along the *z* direction (see Fig. S9) still exhibit no apparent impact to the excitations.

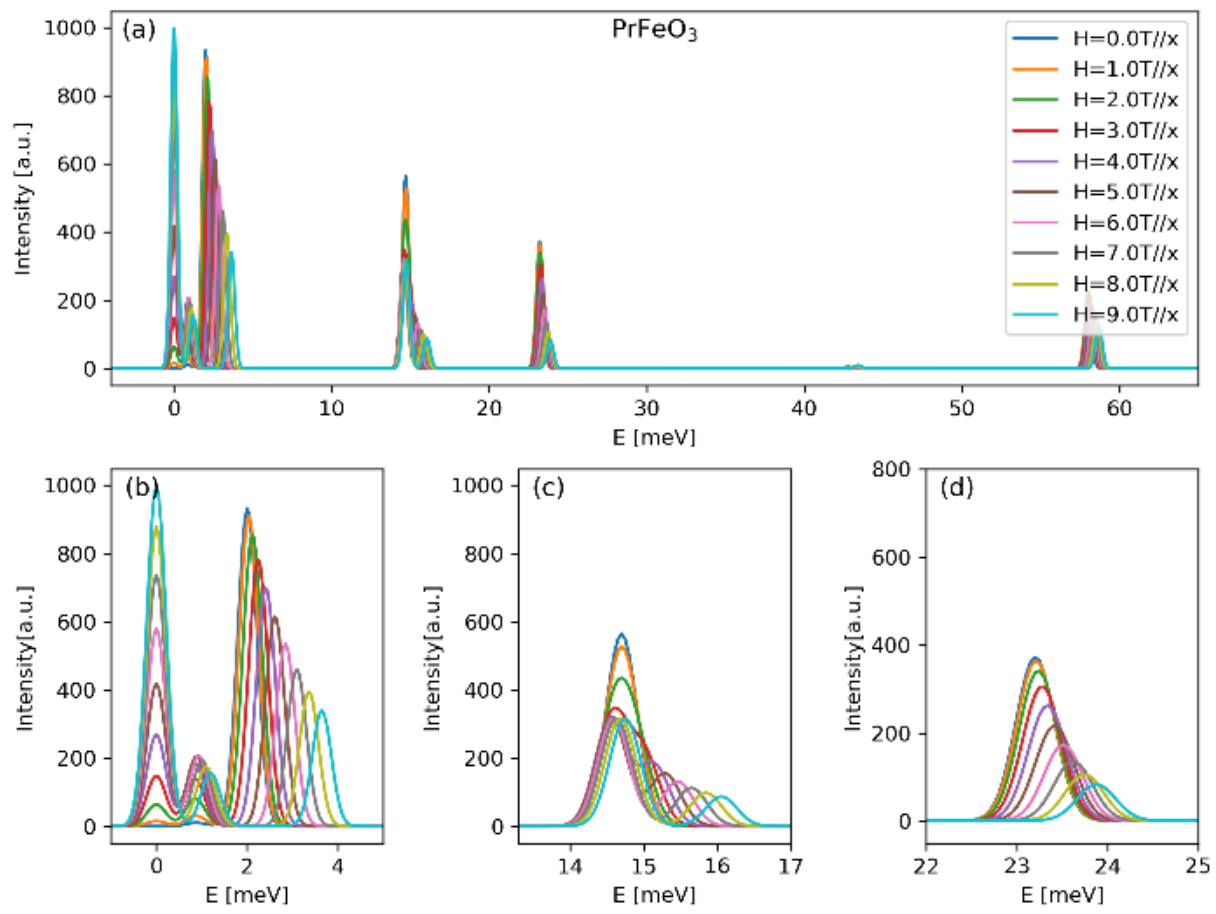

**Fig. S7** (a) CF excitation spectra of $Pr^{3+}$ in $PrFeO_3$ at different external magnetic fields along the *x*-direction. (b), (c), and (d) show zoomed-in views of the first, second, and third CF excitation peaks of $Pr^{3+}$ from (a), respectively.

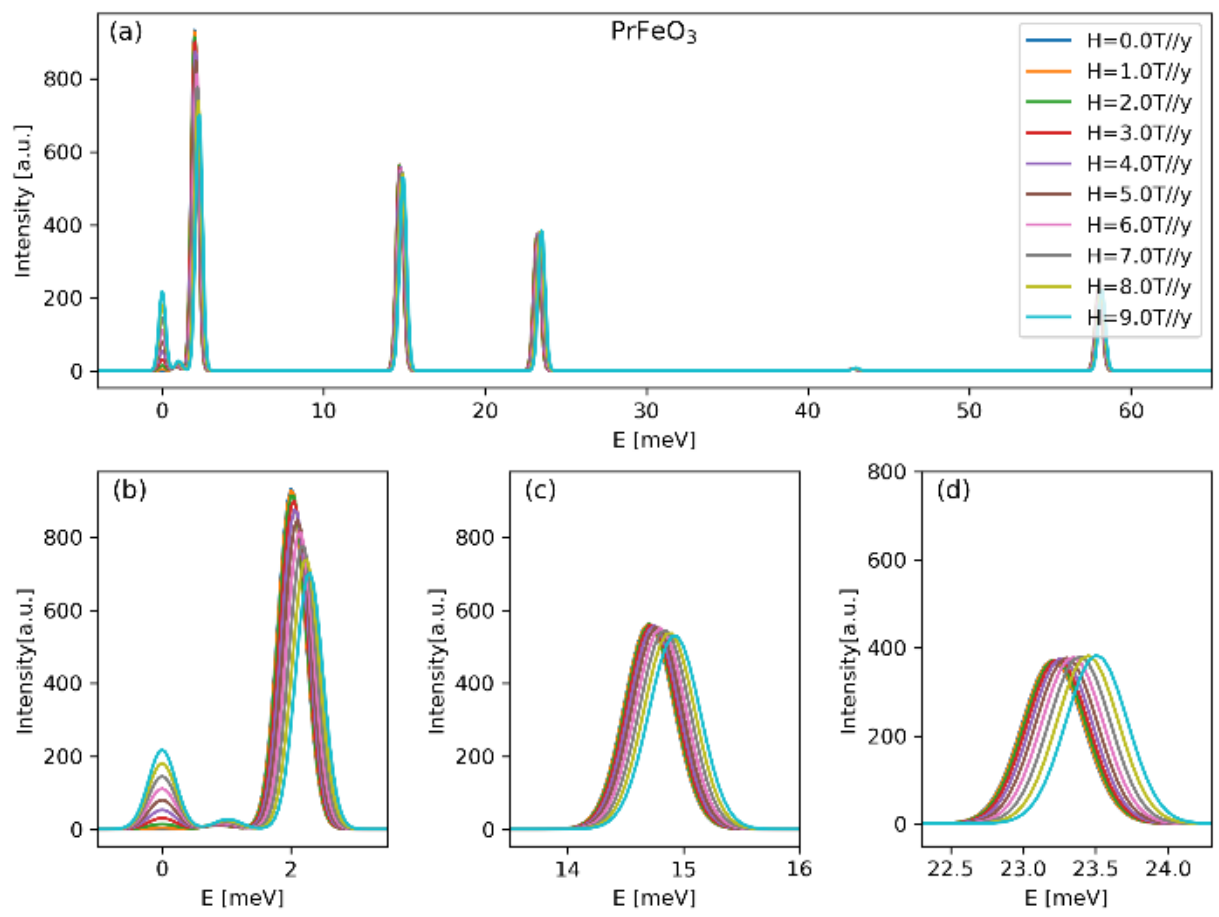

**Fig. S8** (a) CF excitation spectra of $Pr^{3+}$ in $PrFeO_3$ at different external magnetic fields along the *y*-direction. (b), (c), and (d) show zoomed-in views of the first, second, and third CF excitation peaks of $Pr^{3+}$ from (a), respectively.

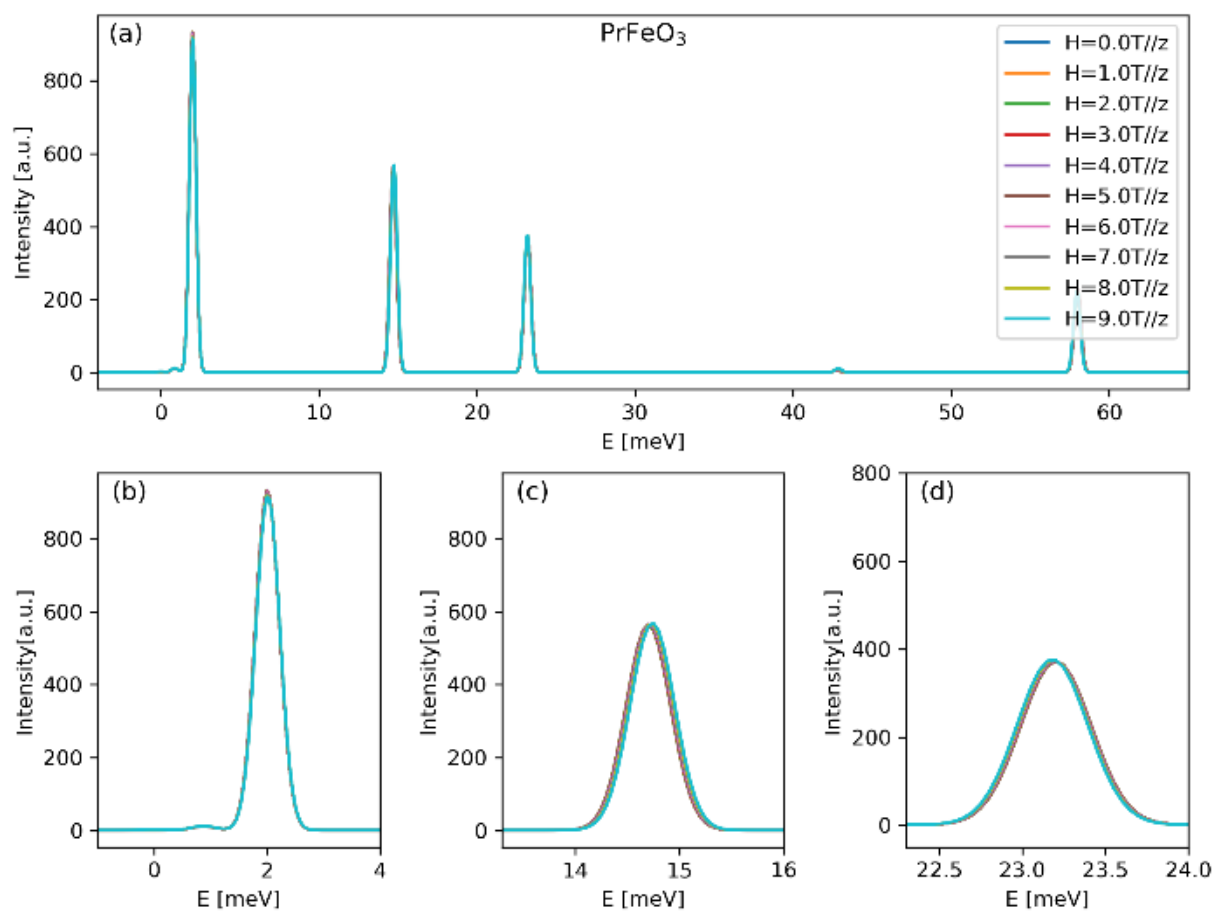

**Fig. S9** (a) CF excitation spectra of $Pr^{3+}$ in $PrFeO_3$ at different external magnetic fields along the *z*-direction. (b), (c), and (d) show zoomed-in views of the first, second, and third CF excitation peaks of $Pr^{3+}$ from (a), respectively.

2.2.2 $HoFeO_3$

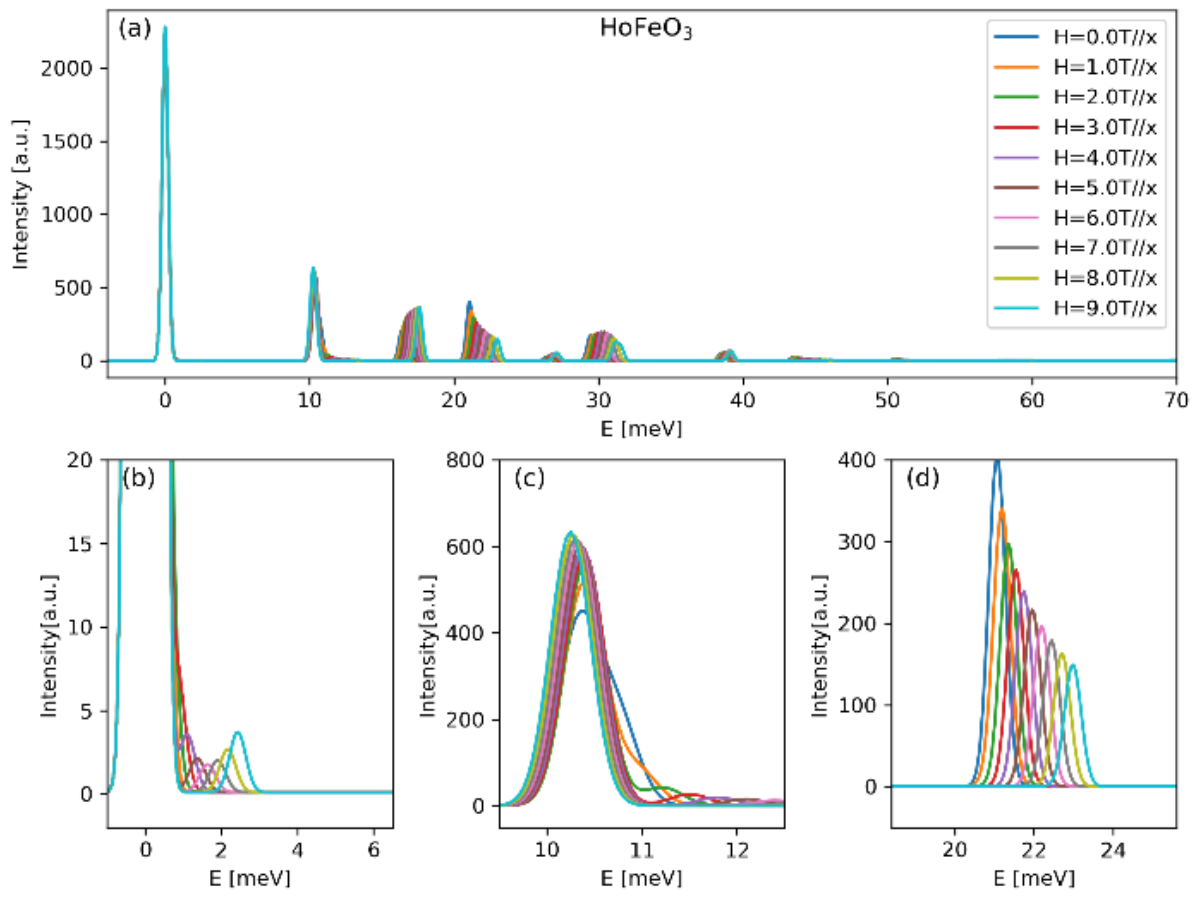

**Fig. S10** (a) CF excitation spectra of $Ho^{3+}$ in $HoFeO_3$ at different external magnetic fields along the *x*-direction. (b), (c), and (d) show zoomed-in views of the first, second, and third CF excitation peaks of $Ho^{3+}$ from (a), respectively.

The influences of the external magnetic fields in $HoFeO_3$ were simulated along the three directions *x*, *y*, and *z*. As shown in Figs. S10, S11, and S12, when the external magnetic fields are applied, the impacts of the fields along the different directions are like what were observed in the internal magnetic fields. However, the effects are much stronger along all three different directions at the same level of the fields. Especially, along the *y* direction, the excited states at 15 meV and 20 meV are significantly shifted by the external fields. The split peak from 10 meV pseudo-doublet have crossed the original 15 meV peak while the 15 meV peak shifts to ~20.5 meV and the 20 meV peak shifts to ~26.5 meV at 9 T. These peak shifts were the largest observed in all the $REFeO_3$ discussed in this work.

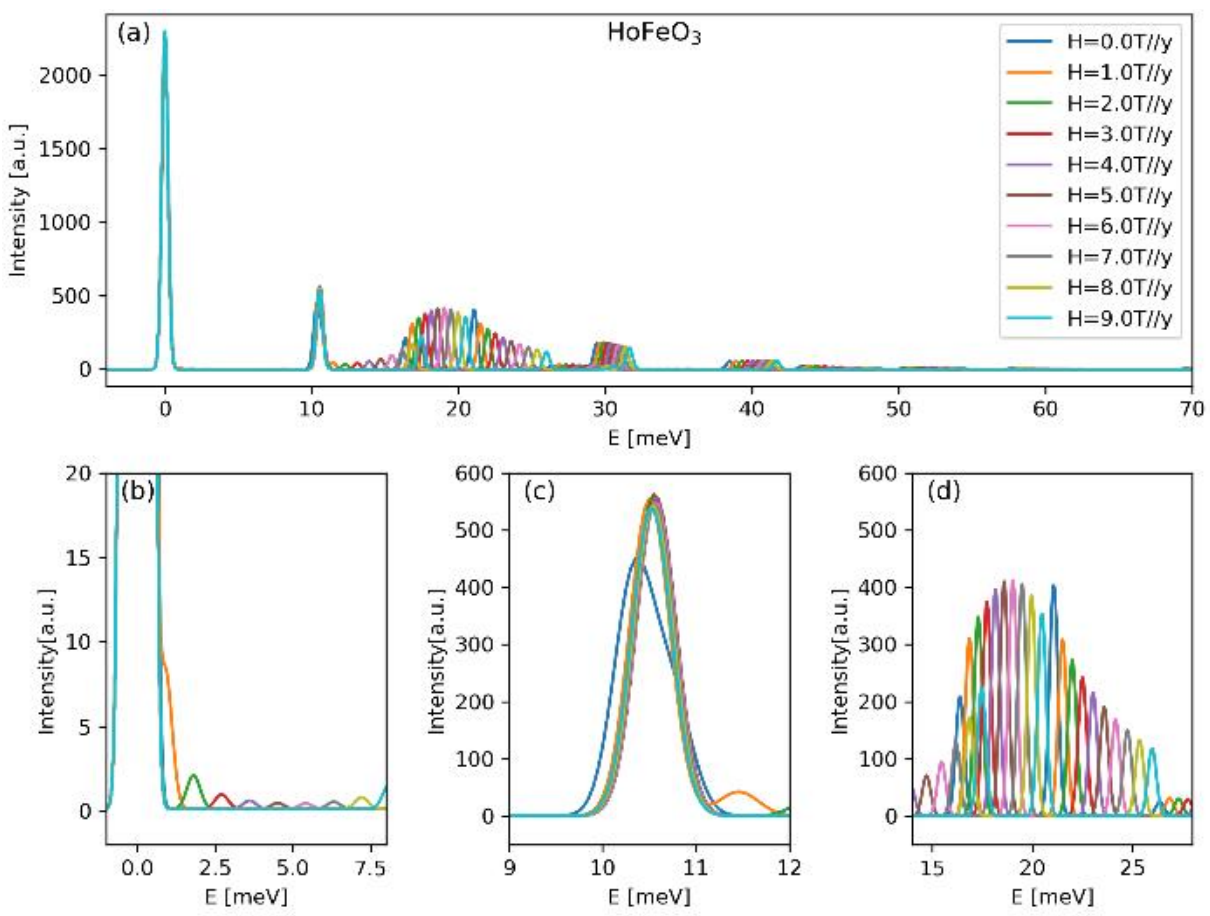

**Fig. S11** (a) CF excitation spectra of $Ho^{3+}$ in $HoFeO_3$ at different external magnetic fields along the *y*-direction. (b), (c), and (d) show zoomed-in views of the first, second, and third CF excitation peaks of $Ho^{3+}$ from (a), respectively.

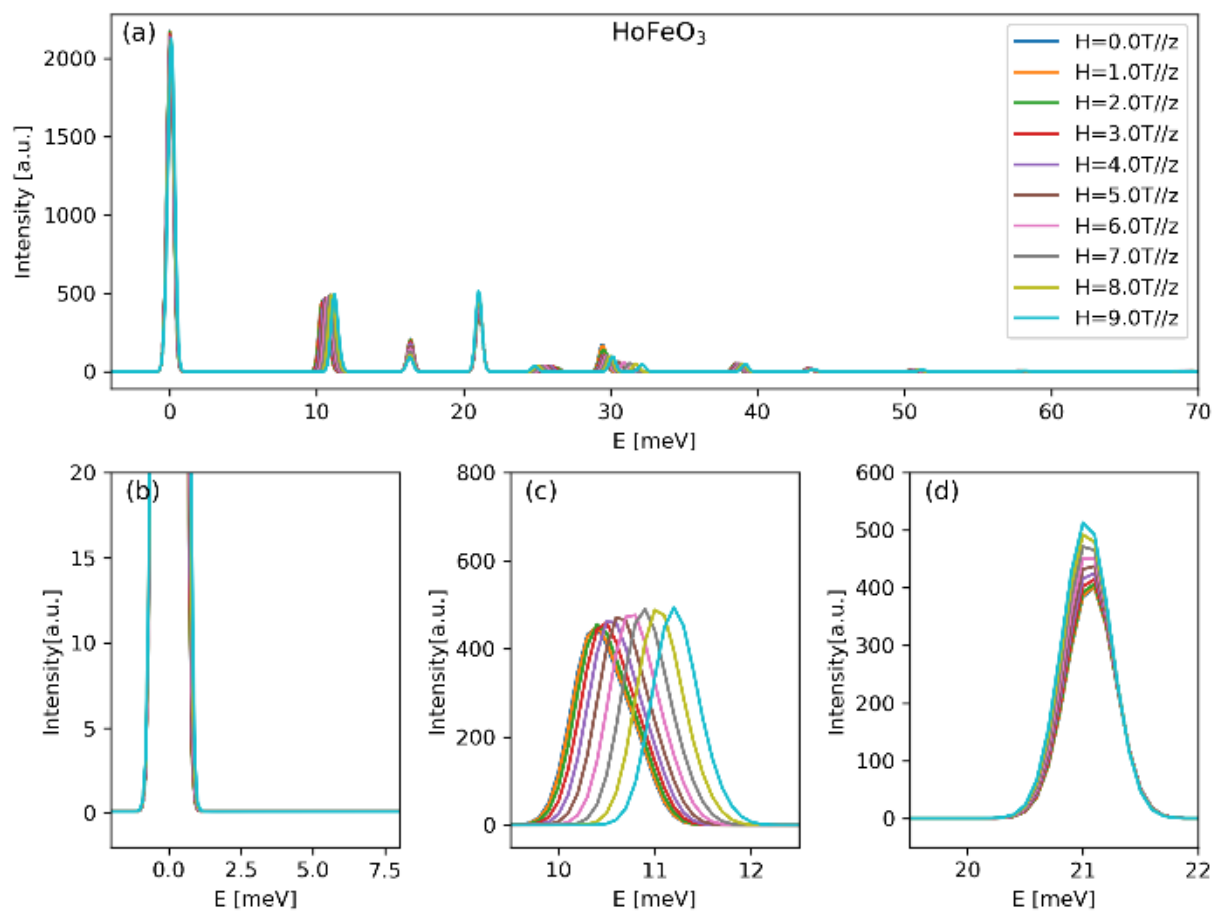

**Fig. S12** (a) CF excitation spectra of $Ho^{3+}$ in $HoFeO_3$ at different external magnetic fields along the *z*-direction. (b), (c), and (d) show zoomed-in views of the first, second, and third CF excitation peaks of $Ho^{3+}$ from (a), respectively.